\documentclass[preprints,article,accept,pdftex,moreauthors]{Definitions/mdpi} 
\firstpage{1} 
\pubvolume{1}
\issuenum{1}
\articlenumber{0}
\pubyear{2025}
\copyrightyear{2025}
\externaleditor{Firstname Lastname} 
\datereceived{ } 
\daterevised{ } 
\dateaccepted{ } 
\datepublished{ } 
\hreflink{https://doi.org/} 

\usepackage{MnSymbol}
\usepackage{amsmath}
\usepackage[symbol]{footmisc}

\DeclareMathOperator*{\argmin}{argmin}

\newcommand{\rs}[1]{{\color{black} #1}}
\newcommand{\rt}[1]{{\color{black} #1}}

\renewcommand\hl[1]{#1} 

\graphicspath{{figures/}}

\Title{\hl{Influence of Boundary} Conditions and Heating Modes on the {Onset of Columnar Convection} in Rotating Spherical Shells}

\TitleCitation{Influence of Boundary Conditions and Heating Modes on the {Onset of Columnar Convection} in Rotating Spherical Shells}

\Author{\hl{William} 
 Seeley $^{\dagger}$ \hl{\orcidD{}}, Francesca Coke $^{\dagger}$ \hl{\orcidC{}}, Radostin D.~Simitev \orcidB{} and Robert J.~Teed *\orcidA{}}

\AuthorNames{William Seeley, Francesca Coke, Radostin D.~Simitev and Robert J.~Teed}

\isAPAStyle{%
       \AuthorCitation{Seeley, W., Coke, F., Simitev, R., \& Teed, R.}
       }{%
        \isChicagoStyle{%
        \AuthorCitation{William Seeley, Francesca Coke, Radostin Simitev and Robert Teed.}
        }{
        \AuthorCitation{\hl{Seeley,} 
 W.; Coke, F.; Simitev, R.D.; Teed, R.J.}
        }
}

\address[1]{%
School of Mathematics and Statistics, University of Glasgow, Glasgow G12 8QQ, UK; 
\linebreak  \hl{qcd511@york.ac.uk (W.S.);
3001175c@student.gla.ac.uk (F.C.);} 
\hl{ radostin.simitev@glasgow.ac.uk (R.D.S.)} 
}

\corres{\hangafter=1 \hangindent=1.05em \hspace{-0.82em}Correspondence: \hl{robert.teed@glasgow.ac.uk}
}

\firstnote{\hangafter=1 \hangindent=1.05em \hspace{-0.82em}The first two authors contributed equally to this work.}  

\abstract{
We investigate the linear onset of thermal convection in rotating
spherical shells with a focus on the influence of mechanical boundary
conditions and thermal driving modes. Using a spectral method, we
determine critical Rayleigh numbers, azimuthal wavenumbers, and
oscillation frequencies over a wide range of Prandtl numbers and shell
aspect ratios at moderate Ekman numbers. We show that the preferred
boundary condition for convective onset depends systematically on both
aspect ratio and Prandtl number: for sufficiently thick 
shells or for
large $\text{Pr}$, the Ekman boundary layer at the outer boundary becomes
destabilising, so that no‑slip boundaries yield a lower $\text{Ra}_c$ than
stress‑free boundaries. Comparing differential and internal heating,
we find that internal heating generally raises $\text{Ra}_c$, shifts the
onset to larger wavenumbers and frequencies, and relocates the
critical column away from the tangent cylinder. 
Mixed boundary conditions with no-slip on the inner boundary
behave similarly to purely stress‑free boundaries,
confirming the dominant influence of the outer surface. These results
demonstrate that boundary conditions and heating mechanisms play a
central role in controlling the onset of convection and should be
carefully considered in models of planetary and stellar interiors. 
}


\begin{document}




\section{Introduction}
\label{sec:intro}


{
\emph{\hl{Topic and context.} 
}
Thermal convection in rotating spherical fluid shells serves as a
fundamental model for understanding large-scale fluid motion in the
interiors of planets and stars. These flows play a critical role in
transporting heat and angular momentum and are widely believed to
underlie the generation of global magnetic fields via dynamo
action~\cite{Busse2015}. Geophysical applications include 
Earth's liquid outer core and its geomagnetic field
\cite{Dormy2025}, while
astrophysical relevance spans convective envelopes in stars such as
the Sun~\cite{Charbonneau2014} and the deep interiors of giant planets such as
Jupiter and Saturn \cite{Jones2011a,Tikoo2022}. 

\emph{\hl{Importance of the primary instability.
}}
Although turbulent convection and self-sustained dynamos operate far
above onset, their basic structure and dominant balances are often
rooted in the nature of the primary instability.
{A wide array of nonlinear regimes are possible, ranging from vacillating oscillations, localised convection, and relaxation oscillations (between convective and zonal flows) (e.g., \cite{tilgner1997,simitev2003}) to turbulence at high levels of supercriticality (e.g.,~\cite{christensen2002,king2013,fan2024})}.
Nevertheless,} the dominant linear
convective modes dictate the preferred spatial and temporal scales of
motion and strongly influence the morphology of the nonlinear state
\cite{simitev2003,busse2005}.
{Additionally,} the distance to onset is frequently used as a reference value to quantify
regimes of nonlinear convection and dynamo action
\cite{christensen2006,Dormy2025}. Consequently, a systematic understanding of
the linear onset problem is essential for interpreting simulations and
extrapolating to astrophysical conditions. 
{In fully nonlinear simulations, convection is found to onset within 1\% of the critical parameter values determined by the linear analysis; e.g.,~see Figure 3 of \cite{Simitev2011}.} {However, the linear mode is not an approximation for nonlinear convection, especially when operating far beyond the onset.} 

\emph{\hl{Asymptotic theories of the convective onset. 
}}
The theoretical framework for the onset of convection in {strongly} rotating spheres has developed along two asymptotic paths in the parameter space of the problem (parameters are defined in Section \ref{sec:methods}).
The classical double limit of a small Ekman number $\text{E} \ll 1$ in
conjunction with a large Prandtl-to-Ekman number ratio $\text{Pr}/\text{E}
\gg 1$ was first studied by Roberts \citep{roberts1968} and Busse
\cite{busse1970} following the pioneering 
formulation of the problem by Chandrasekhar \cite{chandrasekhar1961}. 
These analyses revealed that convection takes the form of strongly
localised, non-axisymmetric columnar rolls centred on a critical
cylindrical radius and with critical Rayleigh number scaling 
$\text{Ra}_c \sim \text{E}^{-4/3}$, consistent with the column width scaling $\sim
\text{E}^{-1/3}$. However, Soward \cite{Soward1977} showed that the presence of a
non-zero radial frequency gradient causes phase mixing, invalidating
the predicted onset condition unless a global criterion is
enforced. Jones et al.~\cite{jones2000} and Dormy et
al.~\cite{dormy2004} resolved this 
issue by analytic continuation into the complex radial plane, removing
the phase mixing and yielding quantitatively accurate predictions for
finite Prandtl number convection in full spheres and spherical shells,
respectively.
A distinct inertial convection regime in the double limit of a small
Ekman number $\text{E} \ll 1$ and a small Prandtl-to-Ekman number ratio
$\text{Pr}/\text{E} \ll 1$  featuring a scaling $\text{Ra}_c \sim \text{E}$ was identified by
Zhang \cite{Zhang1994} and Ardes et al.~\cite{Ardes1997}. A perturbation theory
assuming inertial wave modes sustained by buoyancy against weak
viscous dissipation at the next order of approximation was offered by
Busse \& Simitev \cite{busse2004}. Zhang et al.~\cite{ZHANG2001} showed that interior
viscous dissipation of inertial waves vanishes identically in a full
sphere, which enabled Zhang et al.~\cite{ZHANG2004,ZHANG2007} to propose a unified
asymptotic framework valid in both limits, in which solutions are
sought as a superposition of quasi-geostrophic-inertial-wave modes. 

\emph{\hl{Onset of convection in natural systems. 
}}
Natural systems invariably deviate from the idealised conditions
assumed in the asymptotic theories discussed above. They feature
additional structural and physical complexity \cite{Braginsky1995}
and, of course, finite values of $\text{E}$ and $\text{Pr}$. For example,
{there are significant differences in shell thickness, heterogeneous distribution of
thermal forcing,  material properties, and interface conditions between the convective regions of different planets and stars.}
Earth's outer core is {an iron-nickel alloy thought to be stably stratified at its upper boundary \cite{Dormy2025}, gas giants such
as Jupiter and Saturn exhibit strong internal heating and 
convective layers of metallic or molecular hydrogen, while the convective regions of Uranus and Neptune are likely composed of ionic water ices; see \cite{Busse2015,Soderlund2025} and references within.} Similarly, solar
convection occurs within a shallow shell bounded by the photosphere
and the radiative zone, with heat generated in the core and mixed
thermal and mechanical conditions \cite{Miesch2009}. These deviations can significantly
influence both the structure of convection and the critical parameters
at onset and necessitate numerical instability analyses tailored to
specific interior models, e.g.,~\cite{Garcia2018,Silva2019,barik2024,Morison2024}.

\emph{\hl{Goals of the study. 
}}
Motivated by these considerations, the goal of the present study is to
examine the influence of mechanical boundary conditions
and thermal forcing type on the onset of convection in rotating
spherical shells. In contrast to most previous studies that adopt
one particular model formulation---e.g.,~no-slip boundaries
with purely differential heating---we consider a broader range of 
configurations. Specifically, 
we explore the effects of no-slip, stress-free, and mixed mechanical
boundary conditions, and compare purely internal versus differential
heating models. We take the opportunity to compare and explore these
model variations across a wide range of Prandtl numbers and shell
thicknesses, bridging the gap between thin- and thick-shell regimes. 
In this way, our investigation extends and complements prior work
focused on the roles of Prandtl number~\cite{Ardes1997,simitev2003,Busse2006,fan2024} and radius ratio
\cite{Al-shamali2004,Silva2020,barik2024} by introducing systematic
variation in heating modes 
and boundary 
conditions. Of particular interest are the following questions: 
\begin{enumerate}
\item For which values of $\mathrm{Pr}$ and radius ratio are no-slip boundaries preferred
over stress-free ones, and how does this depend on the choice of heating? 
\item To what extent do mixed boundary conditions interpolate or alter
the behaviour observed in the purely no-slip and purely stress-free
cases? 
\item How does the location of convective onset shift with radius ratio and $\mathrm{Pr}$,
especially within the intermediate-shell range? 
\end{enumerate}
\hl{To} 
 facilitate this exploration, we employ moderate Ekman numbers that
permit extensive parametric coverage while retaining relevance to
{strongly} rotating systems. {Note that such systems are frequently referred to as ``rapidly rotating'' in the geophysical and planetary context but should not be confused with the rapid rotation regime in which} {stars become oblate spheroids}. 
The results of our work offer a refined understanding of
the convective onset problem and provide guidance on model selection
for future investigations of nonlinear convection and magnetic field
generation in spherical geometries.

\section{Models and Methods}
\label{sec:methods}

\subsection{Baseline Linearised Model of Rotating Spherical Thermally-Driven Convection}

\begin{figure}[H]
    \includegraphics[width=0.4\linewidth]{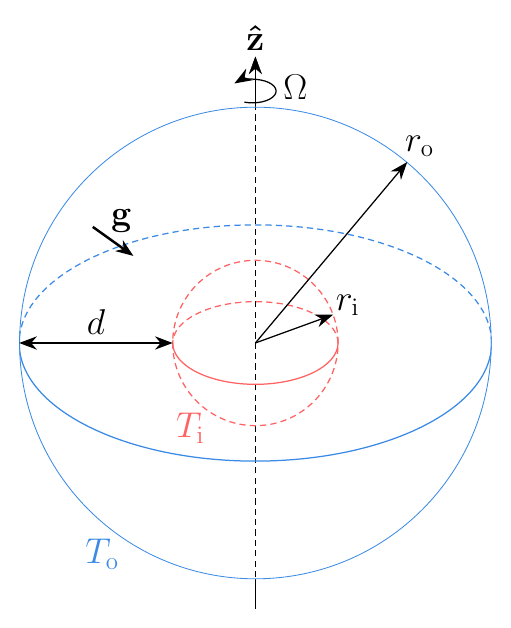}
    \caption{Schematic diagram of the problem. The inner spherical surface at $r=r_{\text{i}}$ is held at temperature $T=T_{\text{i}}$ and the outer spherical surface at $r=r_\text{o}$ is held at temperature $T=T_{\text{o}}<T_{\text{i}}$. The system rotates about the vertical ($z$-direction) with rotation rate, $\Omega$. The gap width is $d=r_{\text{o}}-r_{\text{i}}$ and gravity acts radially, $\mathbf{g}=-gr\mathbf{\hat{r}}$.}
    \label{shellFigure}
\end{figure}

{\emph{\hl{Geometric configuration. 
}}
We consider an incompressible viscous fluid confined within a spherical
shell of inner \hl{radius} 
 $r = r_{\text{i}}$ and outer radius $r =
r_{\text{o}}$. The shell rotates uniformly about the vertical
$\mathbf{\hat{z}}$-axis with constant angular velocity
$\boldsymbol{\Omega} = \Omega \mathbf{\hat{z}}$. {The fluid is subject to a static radial gravitational acceleration,} $-g r
\mathbf{\hat{r}}$, where $\mathbf{\hat{r}}$,
$\boldsymbol{\hat{\theta}}$, and $\boldsymbol{\hat{\varphi}}$ are the
orthogonal unit vectors of the spherical polar coordinate system denoted 
$(r,\theta, \phi)$.
The configuration is illustrated in Figure~\ref{shellFigure}.
}

{\emph{\hl{Thermal driving. 
}}
To model convection driven by thermal gradients, we adopt the
Boussinesq approximation, under which all material properties are
treated as constant, except for the density, which retains a linear
dependence on temperature when contributing to the buoyancy
force. 
We denote the temperature at the inner and outer boundaries by
$T_{\text{i}}$ and $T_{\text{o}}$, respectively, and define the
temperature contrast $\Delta T = T_{\text{i}} - T_{\text{o}} > 0$. Two
standard heating mechanisms are considered as follows. In a pure
``differential heating'' case the temperature values $T_{\text{i}}$ and $T_{\text{o}}$ at
the boundaries are imposed and there is no internal heat source. The static temperature
field $T_0(r)$ then satisfies the Laplace equation $\nabla^2 T_0 =
0$. In a pure ``internal heating'' case, a uniform 
volumetric heat source $S > 0$ is assumed, leading to a static state
satisfying the Poisson equation $\nabla^2 T_0 = -S$. {A constant $S$ is a reasonable approximation, as convective mixing within the fluid shell tends to homogenise the distribution of heat sources \cite{Braginsky1995, davies2011}.} In the latter case,
the
temperature values $T_\text{i}$ and $T_\text{o}$ are not imposed but
determined self-consistently to satisfy $\Delta T = S(r_\text{o}^2-r_\text{i}^2)/6$ by
imposing the regularity condition that the term proportional to $r^{-1}$ in the
corresponding solution vanishes identically. This ensures that no singularity arises when the
same formulation is applied to a full sphere. In summary, the static
temperature field $T_0(r)$ is then found to be
\begin{flalign}
\label{basic temperature state heating cases}
T_0(r)=  
\begin{cases}
-\dfrac{\Delta T(1-\chi)}{d^2(1+\chi)}r^2 + \dfrac{T_\text{i} - \chi^2 T_\text{o}}{(1+\chi)(1-\chi)}& \text{for internal heating,}\vspace{0.5cm}\\
\dfrac{d\chi\Delta T}{(1-\chi)^2}r^{-1} + \dfrac{T_\text{o}-\chi T_\text{i}}{1-\chi}& \text{for differential heating,}
\end{cases}
\end{flalign}
where $d = r_{\text{o}} - r_{\text{i}}$ is the shell thickness and
$\chi = r_{\text{i}} / r_{\text{o}}$ is the radius ratio. These
expressions are the same as those used in \cite{dormy2004}.
The resulting temperature gradients that drive convection can be
written in the compact form
\begin{subequations}
\label{nablaT0}
\begin{gather}
\label{nablaT0beta}
\nabla T_0(r) = -\beta\Lambda(1-\chi)\Lambda(r)\mathbf{\hat{r}},
\end{gather}
or in the alternative form
\begin{gather}
\label{nablaT0DeltaT}
\nabla T_0(r) = -\Delta T d^{-1} \Gamma(\chi)\Lambda(1-\chi)\Lambda(r)\mathbf{\hat{r}},    
\end{gather}
\end{subequations}
with the help of the real-valued functions
\begin{gather*}
    \Lambda(x)= 
    \begin{cases}
        x& \text{for internal heating,}\\ 
        x^{-2}& \text{for differential heating,}
    \end{cases}
~~~~~~~~
    \Gamma(x)= 
    \begin{cases}
        2(1+x)^{-1} & \text{for internal heating,}\\ 
        x & \text{for differential heating,}
    \end{cases}
\end{gather*}
and where $\beta = -|\nabla T_0(r_\text{o})|$ denotes the temperature
gradient at the outer boundary. 
\hl{\mbox{Equations~(\ref{nablaT0}a,b)}} 
 provide a relationship between $\beta$ and
$\Delta T$ (i.e.,~$\beta d = \Delta T\Gamma(\chi)$) which facilitates
conversion between alternative formulations and comparison with other
studies (see \mbox{Section \ref{sec:altform}}).
}

{\emph{\hl{Linearised governing equations and nondimensional parameters. 
}}
To model the onset of convection, we now consider small velocity,
$\mathbf{u}$, and temperature, $\Theta$, perturbations  about the
hydrostatic state of no motion and background temperature given by
Equation \eqref{basic temperature state heating cases}. Using the 
shell thickness $d = r_\text{o} - r_\text{i}$ as the unit of
length, the viscous diffusion time $d^2/\nu$ as the unit of time, and
$\nu\beta d/\kappa$ as the temperature scale, the
non-dimensional linearised governing equations take the \hl{form} 
}
\begin{subequations}
\label{f1eqns}
\begin{gather}
   \label{momentum equation 1}
        \frac{\partial\mathbf{u}}{\partial t} + 2\text{E}^{-1}\mathbf{\hat{z}}\times\mathbf{u} = -\nabla p+\text{Ra}\Theta r\mathbf{\hat{r}} + \nabla^2\mathbf{u},\\
    \label{heat equation 1}
        \text{Pr}\frac{\partial\Theta}{\partial t}=\Lambda(1-\chi)\Lambda(r)\mathbf{u}\cdot\mathbf{\hat{r}}+\nabla^2\Theta,\\
    \label{mass equation 1}
        \nabla\cdot\mathbf{u}=0,
    \end{gather}
\rs{where $p$ is a modified excess pressure, $\nu$ is the kinematic
viscosity, and $\kappa$ is the thermal diffusivity, and the Ekman, Rayleigh, and
Prandtl numbers, defined as 
\begin{gather}
            \text{E} = \frac{\nu}{\Omega d^2}, ~~~~
            \text{Ra} = \frac{\alpha g\beta d^5}{\nu\kappa},~~~~~
            \text{Pr} = \frac{\nu}{\kappa},
            \label{f1params}
\end{gather}
appear as dimensionless parameters.}
\end{subequations}
Note that the factor of $\Lambda(1-\chi)$ within the second term of
Equation~(\ref{heat equation 1}) arises through Equation~(\ref{nablaT0beta}) as
a result of the choice of non-dimensionalisation, combined with
$\beta$ being defined at the outer boundary. If the outer radius
$r_\text{o}$ were instead used as the unit of length this factor would not
appear explicitly (see Section \ref{sec:altform}).  
Alternatively, the temperature scale could be redefined so this factor
appears in the buoyancy term of Equation~(\ref{momentum equation 1}),
either explicitly or within a modified definition of the Rayleigh
number. However, we have chosen to retain the $\Lambda(1-\chi)$ factor
within the heat equation partly so the Rayleigh number takes on its
classical form, free of geometric factors. 

{\emph{\hl{Boundary conditions. 
}}
\hl{Equations} (\ref{f1eqns}a--d)
 are complemented by fixed-temperature and impermeable boundary conditions, expressed as}
\begin{subequations}
  \label{BC}
  \begin{gather}
\Theta = 0, \qquad \mathbf{u} \cdot \mathbf{\hat{r}} = 0,
\end{gather}
on both inner and outer spherical surfaces, $r=r_\text{i}, r_\text{o}$. These are supplemented by
a further mechanical boundary condition, which may be either
no-slip,
\begin{gather}
  \mathbf{u} \cdot \boldsymbol{\hat{\theta}} = \mathbf{u}
  \cdot \boldsymbol{\hat{\varphi}} = 0,
  \label{no-slip boundary condition}
\end{gather}
or stress-free,
\begin{gather}
  \nabla \left( \frac{1}{r} \mathbf{\hat{r}} \times \mathbf{u} \right)
  \cdot \mathbf{\hat{r}} = 0,
  \label{stress-free boundary condition}
\end{gather}
\end{subequations}
\rs{or a ``mixed'' combination of no-slip at the inner boundary and stress-free
at the outer boundary independently. 
{We choose to study only this form of mixed conditions because}
{
convection zones of gaseous giant planets and solar-type stars are confined by open space or atmospheres from above (where modelling by stress-free conditions is suitable); {conversely,} the reverse configuration {of mixed conditions} is not expected to be {physically} relevant.
} 
Throughout the text, a superscript ‘NS’ denotes cases employing purely no-slip conditions, while ‘SF’ refers to cases with purely stress-free boundaries.
}

\subsection{Numerical Method of Solution}

{\emph{\hl{Linear analysis in normal modes. 
}}}
\rs{\hl{Equations} (\ref{f1eqns}a--d) \hl{and Equations} (\ref{BC}a--c) 
define a linear, homogeneous
boundary value problem. 
The system is of second-order with variable
coefficients in space, but only first-order with constant coefficients in time due to the steadiness of the basic state. The solution of this class of problem is a linear superposition (or integral) of separable modes of the form $\hat{\mathbf{q}}(\mathbf{x}) e^{\sigma
t}$, where $\sigma = \gamma + i\omega$, with $\gamma, \omega \in \mathbb{R}$, is a complex growth rate. Substituting this ansatz into the governing equations reduces the problem to a Sturm--Liouville differential eigenvalue problem in space only, $\sigma \hat{\mathbf{q}} = \mathcal{A} [\hat{\mathbf{q}}]$, where $\mathcal{A}$ is a linear differential operator corresponding to the spatial structure and the boundary conditions in \hl{Equations} (\ref{f1eqns}a--d) \hl{and Equations} (\ref{BC}a--c). 
  The spectrum of growth rates $\sigma$ governs the temporal behaviour
of each admissible eigenfunction $\hat{\mathbf{q}}(\mathbf{x})$ of $\mathcal{A}$.}

\rs{\emph{\hl{Numerical discretisation. 
}}}
\rs{For numerical treatment, the eigenfunctions are expanded in a finite set of trial orthogonal functions satisfying the boundary conditions. Projection of the operator equation onto this basis yields a generalised matrix eigenvalue problem,
\begin{flalign}
  \label{eigenvalue problem1}
\mathbf{A} \mathbf{v}  = \sigma \mathbf{M} \mathbf{v},
\end{flalign}
where $\mathbf{v}$ is the vector of expansion coefficients, $\mathbf{M}$ is a mass matrix formed by inner products of basis functions, and matrix $\mathbf{A}$ is a
discretised representation of the operator $\mathcal{A}$. The eigenvalues $\sigma$ of the resulting matrix problem approximate those of the continuous system and determine the real growth rates, $\gamma$, and
oscillation frequencies, $\omega$, of the discrete modes. Various alternative choices of trial functions, truncation rules, and spectral projection methods are available, and here we follow the choices used in \cite{jones2009}. 
We present further details of the discretisation in Appendix \ref{discretisation} and the explicit forms of matrices $\mathbf{A}$ and $\mathbf{M}$ in Appendix \ref{matrices}. Here, we only mention that the trial functions are chosen as Chebyshev polynomials of degree $n$ and spherical harmonics of degree $l$ and order $m$, in radial and angular directions, respectively.}

{\emph{\hl{Critical modes and parameter values. 
}}}
\rs{For any prescribed set of control parameters $(\text{Ra},\text{E},\text{Pr}, \rt{\chi})$ and mode
indices $(n,l,m)$, the generalised eigenvalue problem
\eqref{eigenvalue problem1} is solved using an inverse power method iteration procedure, started from an appropriate initial guess, as described further below. {Other methods for this problem, such as Arnoldi iteration~\cite{gibbons2007}, are available; however, as we are interested in solving for a single eigenvalue, and matrices $\mathbf{A}$ and $\mathbf{B}$ are of a size suitable for the inverse power method, this approach is sufficient {and more numerically efficient}.} The discretisation with truncation rates $(N_\text{x},L)$ leads
to matrices $\mathbf{A}$ and $\mathbf{M}$ of finite size, and the
solution yields a discrete spectrum of eigenpairs
$\{(\sigma_j,\mathbf{v}_j)\}$, one for each retained basis function
indexed by $j=(n,l)$. Because the underlying operator $\mathcal{A}$ is
of Sturm--Liouville type, these eigenvalues can be ordered according to
their real parts, and the mode with the largest real part,
\begin{equation*}
\sigma_{\max} = \operatorname*{arg\,max}_j \Re[\sigma_j],
\end{equation*}
dominates the dynamics and is the relevant one for assessing stability. 
This mode is asymptotically stable if $\gamma<0$, unstable if
$\gamma>0$, and marginally stable at the threshold of convective
instability if $\gamma=0$, where $\gamma = \Re[\sigma_{\max}]$. 
To determine the critical Rayleigh number, we then solve the nonlinear equation
\begin{equation*}
\gamma(\text{Ra},\text{E},\text{Pr},\rt{\chi,}m) = 0
\end{equation*}
for $\text{Ra}$, holding the remaining parameters fixed. This root‑finding
problem is treated by Brent’s method, which yields the critical
marginal value
\begin{equation*}
\text{Ra}_c^{(m)} = \text{Ra}_c^{(m)}(\text{E},\text{Pr},\rt{\chi,}m),
\end{equation*}
together with the associated eigenvalue $\sigma_{\max} = i\omega$, whose
imaginary part $\omega = \Im[\sigma_{\max}]$ gives the oscillation frequency
of the marginal mode. Among all admissible azimuthal wave numbers, $m$,
the critical mode is defined as the one attaining the minimum critical
Rayleigh number, i.e., 
\begin{equation*}
\text{Ra}_c = \min_{m} \text{Ra}_c^{(m)}, \qquad  
m_c = \argmin_{m} \text{Ra}_c^{(m)},
\end{equation*}
so that the global critical parameters are $\text{Ra}_c = \text{Ra}_c^{(m_c)}$, $m_c$, and
$\omega_c = \Im\!\big[\sigma(\text{Ra}_c,\text{E},\text{Pr},\rt{\chi,}m_c)\big]$. 

In the computation of critical curves as functions of control
parameters, we adopt a continuation strategy: the parameters are
varied incrementally, and the previously converged eigenpair is used
as the initial iterate in inverse iteration. The minimisation over $m$
is then restricted to a small neighbourhood of the previously
identified $m_c$, which greatly improves efficiency and
robustness. The detailed implementation of these procedures follows
closely that described by \cite{jones2009}; see
Appendix~\ref{Numerics Appendix}. 
\rt{Note that, in what has been described above, a fixed choice for both boundary conditions 
and heating mode is applied. 
The process is repeated for different combinations of boundary conditions and heating modes.}
The eigenproblem is discretised at a spatial resolution of
$N_\text{x}=L=70$; the inverse power method iterations are
terminated when the residual of the matrix eigenproblem, $\gamma$, falls below a
tolerance of $10^{-5}$, and Brent’s method is terminated when the
residual falls below a tolerance of $10^{-5}$. {Convergence tests {to verify that our} resolution {is sufficient} are shown in Appendix \ref{Convergence Test}.}} 

In addition to determining the critical parameters, it is of interest
to identify the spatial location at which instability first sets
in. The instability typically manifests as columnar structures aligned
with the rotation axis and concentrated near a particular cylindrical
radius $s=s_c$. We estimate $s_c$ by locating the maximum of the
squared azimuthally averaged radial velocity in the equatorial plane, 
\begin{equation*}
s_c = \operatorname*{arg\,max}_{r}\!\left[\int_{0}^{2\pi}\!\left|u_r\left(r,\frac{\pi}{2},\phi\right)\right|\,\mathrm{d}\phi\right],
\end{equation*}
where $u_r = \mathbf{u}\cdot\hat{\mathbf{r}}$. {{The value for $s_c$ is computed from $u_r$ constructed on a uniform grid with $N_r$} points in the radial direction. {Assuming the function is smooth,} the distance between {two adjacent} grid points, {$\Delta r = \frac{1}{N_r-1}$}, {provides the order of the error} since the true critical radius may lie anywhere {between two neighbouring points}. {In our work, we use $N_r=161$}, and hence the error bounds for the critical radius are $s_c \pm \frac{\Delta r}{2} = s_c\pm \frac{1}{320}$.}

\subsection{Alternative Nondimensionalisations}
\label{sec:altform}

\rs{
The nondimensionalisation adopted above is not unique. Different studies have used varying choices of characteristic length (e.g.,~shell
thickness $d$ or outer radius $r_\text{o}$), timescale (e.g.,~viscous diffusion
$d^{2}/\nu$ or rotation rate $\Omega^{-1}$ scale),
temperature scale, and either the total temperature $T$ or the
perturbation $\Theta$. To facilitate comparison with previous work and
to clarify how these choices affect the resulting dimensionless
parameter values, we summarise several alternative
nondimensionalisations. For reference, the nondimensionalisation
defined by \hl{Equations} (\ref{f1eqns}a--d)
will be referred to as \textbf{AN‑1}. 
}

{
\emph{\hl{Alternative nondimensionalisation \textbf{AN‑2}. 
}}
A second common choice is to nondimensionalise using the outer radius
$r_\text{o}$ as the length scale instead of the shell thickness $d =
r_\text{o}-r_\text{i}$. With this choice, the timescale and temperature scale become
$r_\text{o}^2/\nu$ and $\nu\beta r_\text{o}/\kappa$, respectively
(rather than $d^2/\nu$ and $\nu\beta
d/\kappa$, as used in \textbf{AN‑1}). This modification alters the definitions of the Ekman number and the Rayleigh number. The governing equations take the form 
\begin{subequations}
  \label{f2eqns}
  \begin{gather}
    \frac{\partial \mathbf{u}}{\partial t} + 2\text{E}_{r_\text{o}}^{-1}\mathbf{\hat{z}\times u}  = -\nabla p + \text{Ra}_{r_\text{o}}\Theta\mathbf{r} + \nabla^2\mathbf{u},\label{momentum equation 2} \\
    \text{Pr}\frac{\partial \Theta}{\partial t}  = \Lambda(r)\mathbf{u \cdot \hat{r}} + \nabla^2\Theta, \label{heat equation 2} \\
    \nabla \cdot \mathbf{u}  =0,
\end{gather}
with dimensionless parameters
\begin{gather}
  \label{f2params}
  \text{E}_{r_\text{o}}=\frac{\nu}{\Omega r_\text{o}^2}, ~~~~~~
\text{Ra}_{r_\text{o}}=\frac{g\alpha\beta r_\text{o}^5}{\nu \kappa}.
\end{gather}
\end{subequations}

\hl{We} 
 refer to the system given by \hl{Equations} (\ref{f2eqns}a--d) 
as nondimensionalisation \textbf{AN‑2}. To match exactly the formulation
used by Dormy et al.~\cite{dormy2004}, the factor of 2 in the Coriolis term
must also be absorbed into the definition of the Ekman number. 
The Rayleigh numbers defined in \eqref{f1params} and \eqref{f2params}
differ only through the length scale appearing in their definition. As
a consequence, the factor $\Lambda(1-\chi)$ in Equation~\eqref{heat
  equation 1} is absent in \textbf{AN‑2}. This difference arises
because the temperature scale in each nondimensionalisation is set by
the temperature gradient at $r_\text{o}$, but the gradient is
evaluated relative to a different length scale in each case. 
}

{\emph{\hl{Alternative nondimensionalisation \textbf{AN‑3}. 
}}
A third option retains the shell thickness $d$ as the characteristic
length scale but uses $\nu\,\Delta T/\kappa$ rather than $\nu\,\beta
d/\kappa$ as the temperature scale. This change modifies the governing
equations and the definition of the Rayleigh number to
\begin{subequations}
\label{f3eqns}
\begin{gather}
    \frac{\partial \mathbf{u}}{\partial t} + 2\text{E}^{-1}\mathbf{\hat{z}\times u}  = -\nabla p + \text{Ra}_{\Delta T}\Theta\mathbf{\hat{r}} + \nabla^2 \mathbf{u}, \label{momentum equation 3}\\
    \text{Pr}\frac{\partial \Theta}{\partial t} = \Gamma(\chi)\Lambda(1-\chi)\Lambda(r)\mathbf{u}\cdot\mathbf{\hat{r}} + \nabla^2\Theta, \label{heat equation 3}\\
    \nabla \cdot \mathbf{u}  =0,
\end{gather}
where
\begin{gather}
\label{f3params}
\text{Ra}_{\Delta T}=\frac{\alpha g \Delta T d^4}{\nu\kappa}.
\end{gather}
\end{subequations}

We refer to the system given by \hl{Equations} (\ref{f3eqns}a--d) 
as nondimensionalisation
\textbf{AN‑3}. Under this choice of temperature scale, the form of
$\nabla T_0$ from Equation (\ref{nablaT0DeltaT}) is implemented,
introducing the additional factor $\Gamma(\chi)$ in the source term of
Equation \eqref{heat equation 3}. As in \textbf{AN‑1}, the factor
$\Gamma(\chi)\,\Lambda(1-\chi)$ may instead be placed in the buoyancy
term or absorbed into the definition of the Rayleigh number, provided
the temperature perturbation is consistently redefined.
}

{\emph{\hl{Alternative nondimensionalisation \textbf{AN‑4}. 
}}
A fourth option, frequently adopted in earlier studies, is to use $\Delta T$ itself (rather than $\nu\Delta T/\kappa = \text{Pr}\,\Delta T$) as the temperature scale and to define the Rayleigh number using $g_\text{o}$ (the gravitational acceleration at the outer boundary) instead of $g = g_\text{o}/r_\text{o}$. With these adjustments relative to \textbf{AN‑3}, the governing equations take the form
\begin{subequations}
\label{f4eqns}
\begin{gather}
    \frac{\partial \mathbf{u}}{\partial t} + 2\text{E}^{-1}\mathbf{\hat{z}\times u}  = -\nabla p + \text{Pr}^{-1}\text{Ra}_{\Delta Tg_\text{o}}(1-\chi)\Theta\mathbf{\hat{r}} + \nabla^2 \mathbf{u},     \label{momentum equation 4}\\
    \frac{\partial \Theta}{\partial t} = \Gamma(\chi)\Lambda(1-\chi)\Lambda(r)\mathbf{u}\cdot\mathbf{\hat{r}} + \text{Pr}^{-1}\nabla^2\Theta,     \label{heat equation 4}\\
    \nabla \cdot \mathbf{u}  =0,
\end{gather}
where
\begin{gather}
\label{f4params}
\text{Ra}_{\Delta Tg_\text{o}}=\frac{\alpha g_\text{o} \Delta T d^3}{\nu\kappa}.
\end{gather}
\end{subequations}

We refer to the system given by \hl{Equations} (\ref{f4eqns}a--d) 
as nondimensionalisation
\textbf{AN‑4}. The change of temperature scale simply redistributes
factors of $\text{Pr}$ within the equations and does not require any
conversion of control parameters. The modification of gravity in the
Rayleigh number introduces an additional factor of $1-\chi$ in the buoyancy term, which in
some studies appears, equivalently, as $r_\text{o}^{-1}$. 
}
  
The nondimensional scales of \textbf{AN-4} are commonly used in previous studies (e.g.,~\citep{Al-shamali2004}), including many spherical dynamo studies (e.g.,~\citep{christensen2006,teed2023}). However, such studies are routinely posed in terms of the total temperature $T=T_0+\Theta$ (rather than $\Theta$). Then, in \mbox{Equation~(\ref{momentum equation 4})}, $\Theta$ is replaced by $T$ and the pressure perturbation is replaced by the total pressure. Furthermore, Equation~(\ref{heat equation 4}) is replaced by
\begin{equation}
\label{heat equation 4T}
    \frac{\partial T}{\partial t} + \mathbf{u}\cdot\nabla T_0  = \text{Pr}^{-1}\nabla^2T + S,
\end{equation}
for a (potentially non-zero) source term, $S$, satisfying $\text{Pr}^{-1}\nabla^2T_0=-S$. The explicit geometrical factors of the second term of Equation~(\ref{heat equation 4}) are then housed implicitly within $\nabla T_0$ in Equation~(\ref{heat equation 4T}).
Some studies (e.g.,~\cite{barik2024, fan2024}) also use an alternative timescale of $\Omega^{-1}$ (rather than the viscous timescale, $d^2/\nu$), a choice that only positions factors of $\text{E}$ differently within the equations.
These final changes do not fundamentally change the formulation, do not lead to conversions of control parameters, and, for that reason, we do not present them as an additional nondimensionalisation.

{\emph{\hl{Conversions. 
}}
In each of the four nondimensionalisations, given by
Equations (\ref{f1eqns}a--d), (\ref{f2eqns}a--d), (\ref{f3eqns}a--d), and (\ref{f4eqns}a--d), we have, for clarity, retained the same symbols for the nondimensional velocity, temperature, and pressure fields. It should be noted, however, that these quantities are expressed in different nondimensional units in each nondimensionalisation. When comparing results across the different nondimensionalisations, appropriate conversions are obtained by rescaling the Rayleigh number, as listed in Table \ref{Table: Conversions} (and, for \textbf{AN-2} only, the Ekman number via $\text{E}_{r_\text{o}} = (1-\chi)^2\text{E}$ ). Note also that any conversion that involves a change in timescale also requires a suitable rescaling of the frequency, $\omega$.
}
\begin{table}[H] 
\caption{\textls[-15]{Conversions for the Rayleigh number between
  nondimensionalisations. The Rayleigh number of \textbf{AN-1} is
  $\text{Ra}$ and the table provides the equivalent value for the
  other  nondimensionalisations.}  \label{Table: Conversions}}
\begin{tabularx}{\textwidth}{CCC}
\toprule
\textbf{AN-2}	& \textbf{AN-3} & \textbf{AN-4}\\
\midrule
$\text{Ra}_{r_\text{o}} = (1-\chi)^{-5}\text{Ra}$ & $\text{Ra}_{\Delta T} = [\Gamma(\chi)]^{-1} \text{Ra}$ & $\text{Ra}_{\Delta Tg_\text{o}} = (1-\chi)^{-1}\text{Ra}_{\Delta T}$\\
\bottomrule
\end{tabularx}
\end{table}

{\emph{\hl{Further remarks. 
}}}
The various nondimensionalisations may have potentially unintended consequences
when interpreting the model and its results. Much of the literature on
convection and dynamo models in spherical shells in recent years has
lent towards using $d$ as the lengthscale of nondimensionalisation,
typically with a fixed value of $\chi$ appropriate to the planet under
consideration (e.g.,~$\chi=0.35$ for Earth). Using the gap width has
its merits when $\chi$ is fixed since it aids comparison with local
models (e.g.,~models in a plane layer with a fixed distance between top
and bottom plates). However, it is not ideally suited to studies where
$\chi$ varies. For example, the volume of the shell, $V_s$, is an
increasing function of $\chi$ when $d$ is used as the lengthscale for
nondimensionalisation since $V_s=4\pi(1-\chi^3)/3(1-\chi)^3$. This
may seem counter-intuitive since a growing inner core would be
expected to lead to a smaller shell volume. In fact, the volume of the
whole spherical system defined by the sphere at $r=r_\text{o}$ is
enlarging due to the choice of units fixed with $d$, which more than
compensates for the volume `lost' as a result of increasing
$\chi$. The trend is reversed if $r_\text{o}$ is used as the
lengthscale for nondimensionalistaion since then $V_s =
4\pi(1-\chi^3)/3$ and the volume of the whole spherical system (defined by
the sphere at $r=r_\text{o}$) is fixed. This results in the shell
volume shrinking as the inner core grows, which is more intuitive,
especially since planets such as Earth are not themselves swelling!

As we shall see in Section \ref{ref:formulation}, in a similar manner to the shell volume, the choice of nondimensionalisation can affect the trend of the critical Rayleigh number when varying $\chi$, via the differing definitions of Ra seen in \textbf{AN-1}-\textbf{AN-4}. This, in turn, can lead to a different physical interpretation of the results depending on which nondimensionalisation is used, and, as such, care must be taken when determining the ease of convective onset in spheres of
different aspect ratios.

\section{Results}

We find the critical values of the Rayleigh number, wavenumber, and frequency for the onset of convection for a variety of values of input parameters and physical set-ups. While previous studies have explored more extreme parameter regimes with low values of Ekman numbers \cite{barik2024}, low and infinite values of Prandtl numbers \cite{net2008, zhang1991}, and higher and lower radius ratios \cite{Al-shamali2004, barik2024}, we restrict our study to more moderate parameters in order to allocate computation resources to focus on a wider set of boundary conditions and heating types. Specifically, we conduct our study with $\text{E}\in\{10^{-4}, 3\times10^{-5}, 10^{-5}\}$, $\text{Pr}\in[0.1,100]$, $\chi\in[0.2,0.8]$, implementing either differential or internal heating, and using either no-slip, stress-free, or mixed mechanical boundary conditions {(as described in Section \ref{sec:methods})}. 
In the asymptotic limit, $\text{E}\to0$, we expect to find that $\text{Ra}_c$ scales with $\text{E}^{-4/3}$, $m_c$ with $\text{E}^{-1/3}$, and $|\omega_c|$ with $\text{E}^{-2/3}$. These predicted scalings are approached in our results, as E is decreased. We also confirm a subset of our results by converting and comparing with previous studies~\citep{fan2024,christensen2006,Al-shamali2004,barik2024}.

\subsection{Dependence on the Prandtl Number}
\label{sec:Pr}

All results presented in this subsection are for a fixed radius ratio of $\chi=0.35$, the value appropriate for the present-day Earth's core. {The lowest Prandtl number we consider is constrained by the regime boundaries of columnar convection. For the Ekman numbers studied, inertial convection is preferred when $\text{Pr}<0.1$ \cite{Ardes1997, busse2004, net2008}.} We begin with a brief discussion of the case with no-slip boundary conditions and differential heating, since this allows us to readily verify our results against previous studies before introducing new~results.

\begin{figure}[H]
\centering
\captionsetup[subfloat]{position=bottom}
\subfloat[\centering\label{DNSNS PR Ra}]{\includegraphics[width=0.32\linewidth]{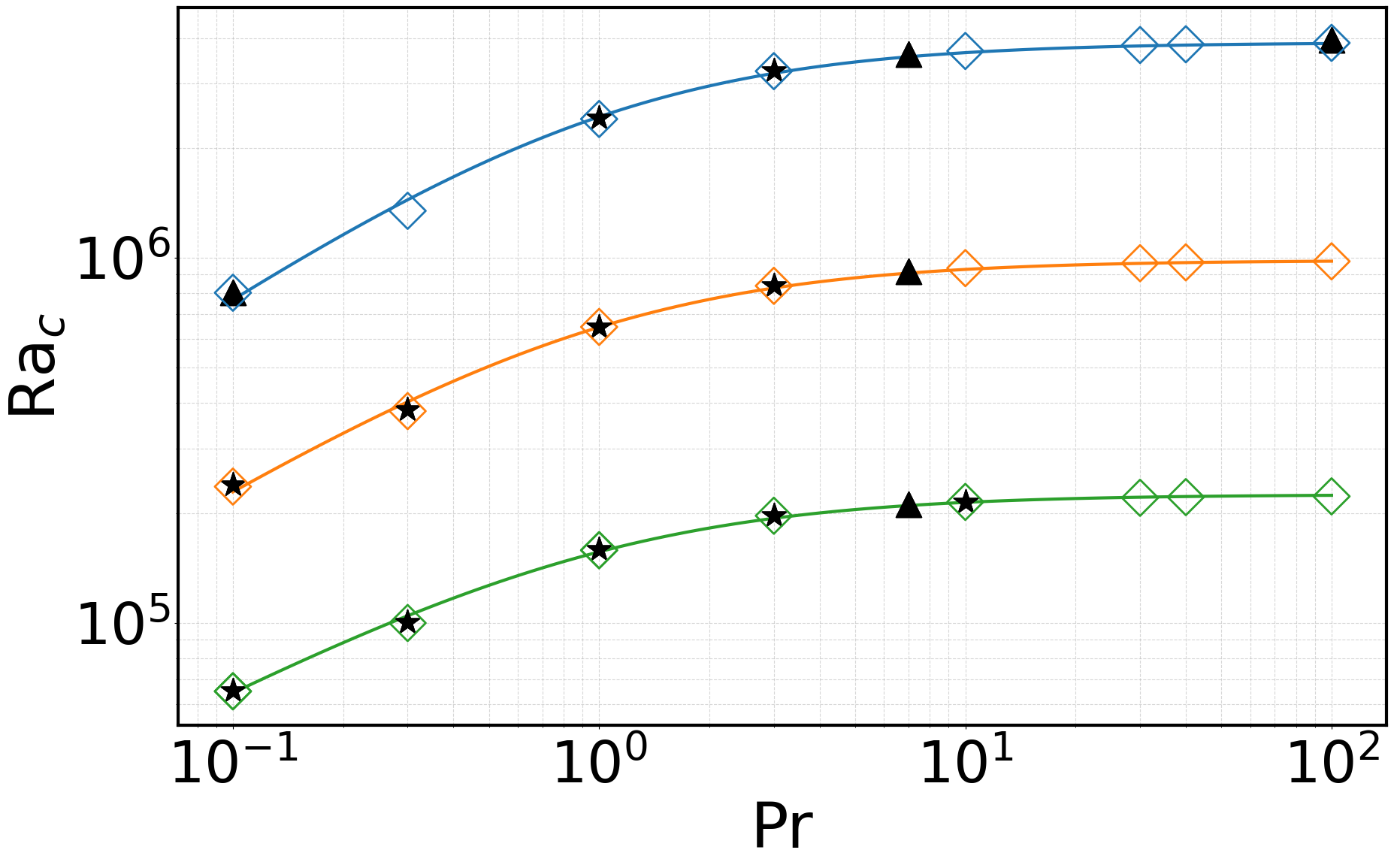}}
\subfloat[\centering\label{DNSNS PR w}]{\includegraphics[width=0.32\linewidth]{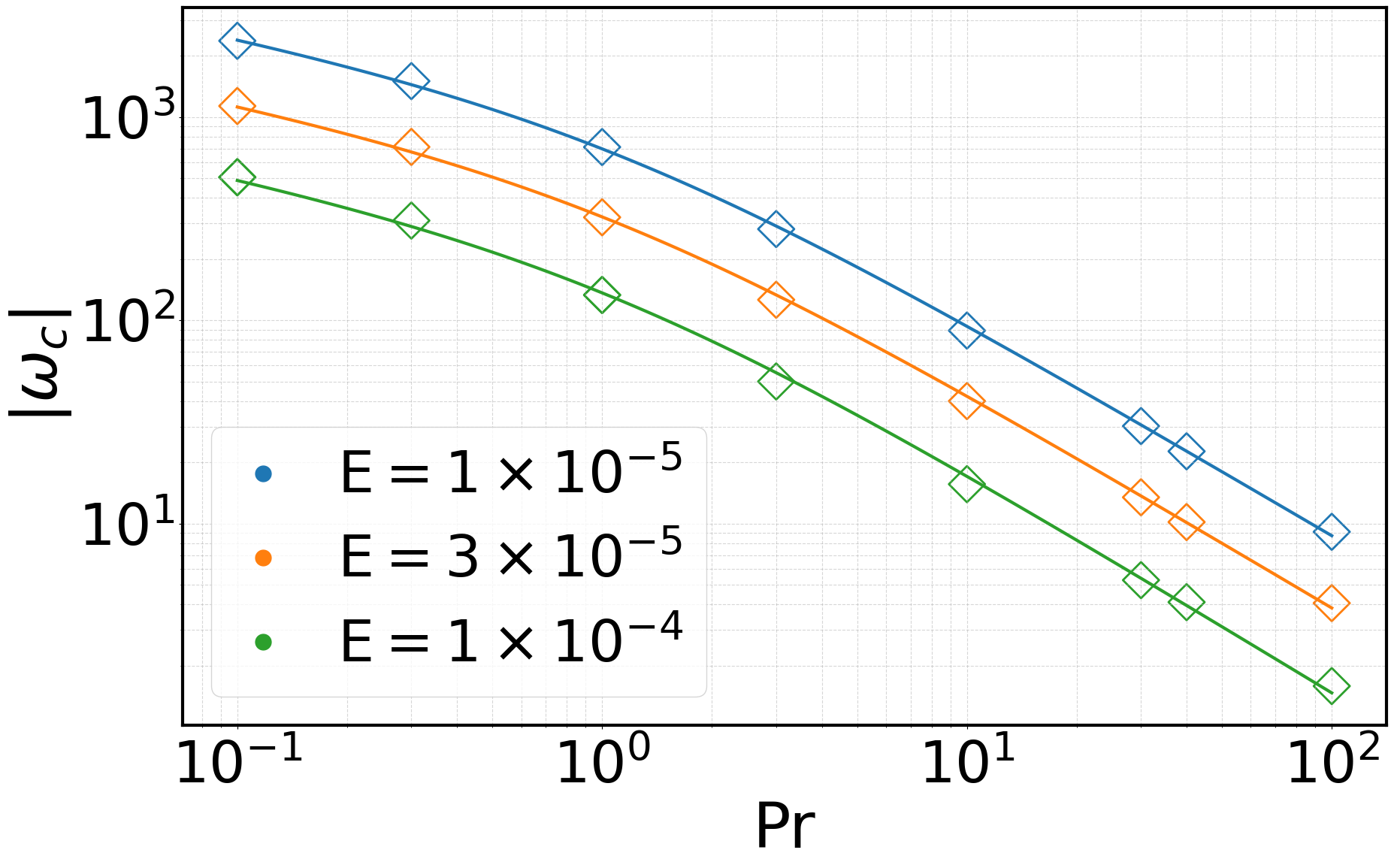}}
\subfloat[\centering\label{DNSNS PR m}]{\includegraphics[width=0.32\linewidth]{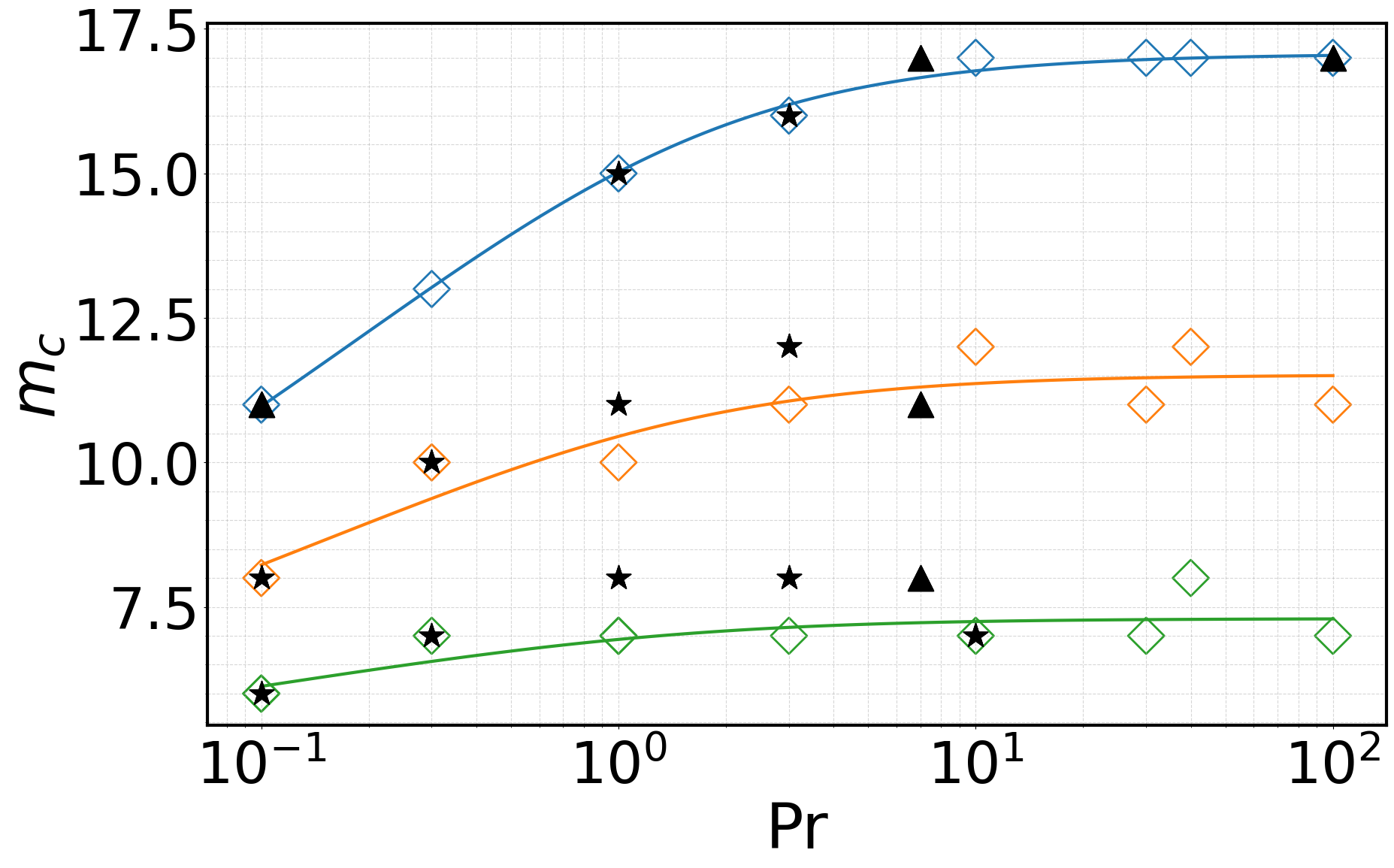}}
\caption{\hl{Dependence} 
 of critical values on the Prandtl number, Pr, for several values of the Ekman number, E, in the case of differential heating with purely no‑slip boundary conditions and radius ratio $\chi=0.35$.
(\textbf{a}) Critical Rayleigh number, (\textbf{b}) critical
  frequency, (\textbf{c}) critical azimuthal wavenumber.
  Critical values found by Christensen and Aubert ($\bigstar$) 
  \cite{christensen2006} and Fan et al ($\blacktriangle$)
    \cite{fan2024} are also shown.
    The best-fit curves are given by Equation (\ref{Pr scaling}) with fitting parameters
    given in Table \ref{Pr fitting exponents}.
        }
\label{DNSNS PR}
\end{figure}

{\emph{\hl{Critical values. 
}}} As previously found, there is a significant effect on the critical values associated with the onset of convection when varying the Prandtl number \cite{zhang1993,simitev2003,fan2024}. Figure~\ref{DNSNS PR}a demonstrates that increasing $\text{Pr}$ leads to a larger critical Rayleigh number and therefore stabilises the system to convective instability. At low Prandtl numbers ($\text{Pr}\lesssim1.0$), the onset of convection is far more sensitive to variations in $\text{Pr}$ than it is at moderate or high $\text{Pr}$. Associated with the increase in critical Rayleigh number as Pr is increased, there is a reduction in the speed of the thermal Rossby waves (Figure~\ref{DNSNS PR}b) and an increase in the number of modes at onset (Figure~\ref{DNSNS PR}c). 

{\emph{\hl{Scaling laws. 
}}} To interpolate the best-fit curves of Figure \ref{DNSNS PR}, we use functions based on the scaling laws found in \cite{simitev2003}. When varying $\text{Pr}$ (with fixed E and $\chi$), we make use of the following three equations:
\begin{subequations}
\label{Pr scaling}
\begin{gather}
\label{Ra Pr scaling}
    \text{Ra}_c=a_1\bigg(\frac{2\text{Pr}}{1+\text{Pr}}\bigg)^{b_1}, \\
\label{m Pr scaling}
    m_c=a_2\bigg(\frac{2\text{Pr}}{1+\text{Pr}}\bigg)^{b_2},\\
\label{omega Pr scaling}
    |\omega_c|=a_3\bigg(\frac{4}{\text{Pr}(1+\text{Pr})^2}\bigg)^{b_3},
\end{gather}
\end{subequations}
where the fitting parameters $a_i$ and $b_i$ (for $i=1,2,3$) are given in Table~\ref{Pr fitting exponents}. As $\text{E}$ decreases, the amplitude $a_i$ grows in accordance with the asymptotic scalings, $\text{Ra}_c\sim \text{E}^{-4/3}$, $m_c\sim \text{E}^{-1/3}$, and $\omega_c\sim \text{E}^{-2/3}$, while the fitting exponents $b_i$ tend to their theoretical values, $b_1\to4/3$, $b_2\to1/3$, and $b_3\to1/3$ as $\text{E}\to0$ \cite{simitev2003}. 

{\emph{\hl{Comparison with previous studies. 
}}} Our results are also validated in Figure~\ref{DNSNS PR}, noting that the critical Rayleigh numbers we find agree (after suitable conversions; see Section \ref{sec:altform}) with previous investigations \cite{christensen2006, fan2024}. There are some small discrepancies between values of $m_c$, which may arise due to differences in methods and the small differences in $\text{Ra}_c$ between values of the discrete variable, $m$. 
Although Christensen and Aubert~\cite{christensen2006} and Fan et al.~\cite{fan2024} did not give explicit values for the critical frequencies, $\omega_c$, we find our results to be consistent with previously found asymptotic scaling laws \cite{simitev2003}.

\begin{table}[H]
 \caption{Fitting parameters $a_i$ and $b_i$ for the critical value best-fit curves from Figure~\ref{DNSNS PR} (for a system with $\chi=0.35$, differential heating, and purely no-slip boundaries). $i=1$ refers to Equation (\ref{Ra Pr scaling}) for $\text{Ra}_c$, $i=2$ refers to Equation (\ref{m Pr scaling}) for $m_c$, and $i=3$ refers to Equation (\ref{omega Pr scaling}) for $\omega_c$.}
    \label{Pr fitting exponents}

\begin{tabularx}{\textwidth}{CcCCCcC}
\toprule
        \textbf{$\text{E}$} & \boldmath{$a_1$} & \boldmath{$b_1$} 
        & \boldmath{$a_2$} & \boldmath{$b_2$} 
        & \boldmath{$a_3$} & \boldmath{$b_3$} \\
        \midrule
        $10^{-4}$  &  $1.556 \times10^5$ &$0.522$
        & $6.938$ &$0.073$ 
        & $1.365\times10^2$ &$0.364$  \\
        \midrule
        $3\times10^{-5}$ &$6.459\times10^5$ & $0.609$ 
        & $10.450$ & $0.140$ 
        & $3.230\times10^2$ & $0.356$  \\
        \midrule
        $10^{-5}$  &  $2.435\times10^6$ & $0.677$  
        & $15.024$ & $0.184$ 
        & $6.981\times10^2$ & $0.352$\\
        \bottomrule
    \end{tabularx}

\end{table}

\begin{figure}[H]
\captionsetup[subfloat]{position=bottom}
\subfloat[\centering\label{Mode Plot DNSNS 0.1}]{\includegraphics[width=0.32\linewidth]{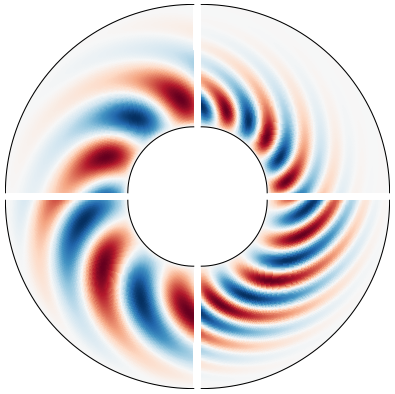}}
\subfloat[\centering\label{Mode Plot DNSNS 1.0}]{\includegraphics[width=0.32\linewidth]{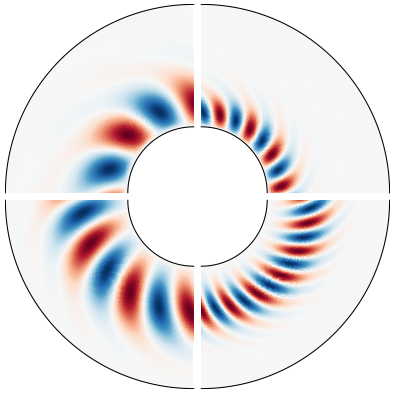}}
\subfloat[\centering\label{Mode Plot DNSNS 100.0}]{\includegraphics[width=0.32\linewidth]{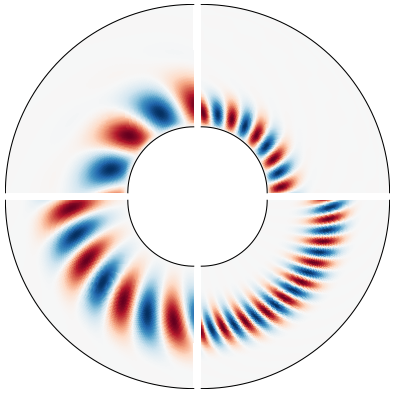}}
\caption{
\hl{The} 
 radial component of the velocity field at onset plotted in an
  equatorial slice for purely no-slip boundary conditions, $\chi=0.35$, differential heating (top half), internal heating (bottom half), 
  $\text{E}=10^{-4}$ (left half of each panel), $\text{E}=10^{-5}$ (right half of each panel), and for (\textbf{a}) $\text{Pr}=0.1$, (\textbf{b}) $\text{Pr}=1.0$, and (\textbf{c}) $\text{Pr}=100.0$. \hl{Red and blue represent positive and negative values, respectively.} \label{Mode Plots DNSNS}}
\end{figure}

{\emph{\hl{Flow patterns. 
}}} The form of the convection columns in the equatorial plane is shown in Figure~\ref{Mode Plots DNSNS}. In the differentially heated case (top half of each plot), the modes are adjacent to the inner boundary as the temperature gradient is largest on the tangent circle. This is in contrast with the internally heated case (bottom half of each plot), which will be discussed further in Section \ref{sec:Prheating}. The plots highlight the dependence of the size of the flow structures on both Ekman number (compare left and right of each plot) and Prandtl number (shown in individual plots). In all cases, as the Prandtl number decreases, the convection columns become increasingly radially elongated, accentuating their spiralling structure.

We now look to further the investigations of \cite{christensen2006,fan2024} by examining the effect of different heating types and mechanical boundary conditions. Specifically, we aim to investigate how internal heating and stress-free boundary conditions influence the dependence of the critical Rayleigh number on $\text{Pr}$ and compare with the differential heated, purely no-slip boundaries case discussed hitherto.

\subsubsection{Effect of Mechanical Boundary Conditions}
\label{sec:PrBCs}

\begin{figure}[H]
\captionsetup[subfloat]{position=bottom}
\subfloat[\centering]{\includegraphics[width=0.487\linewidth]{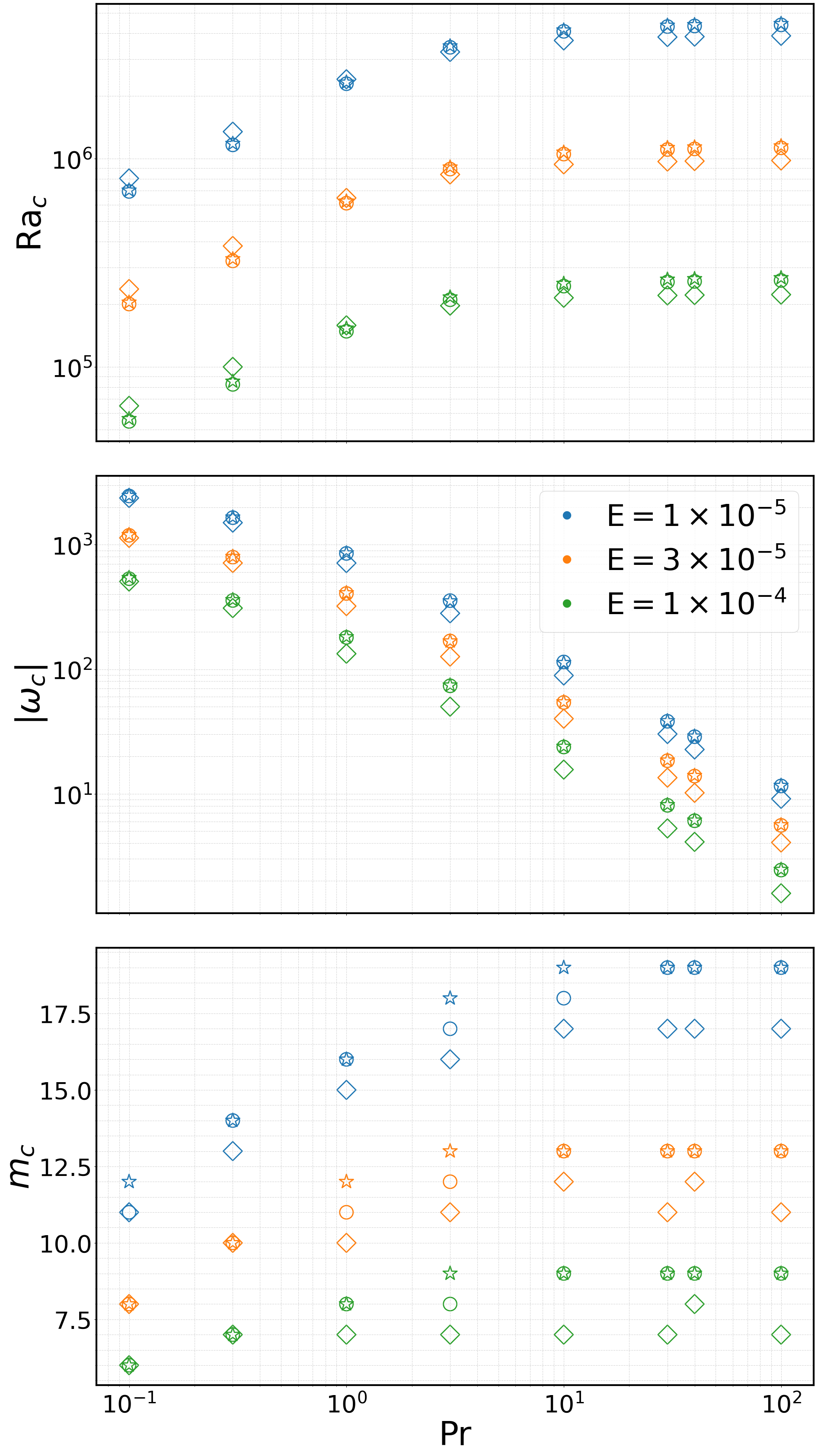}\label{DBCs}}
\subfloat[\centering]{\includegraphics[width=0.48\linewidth]{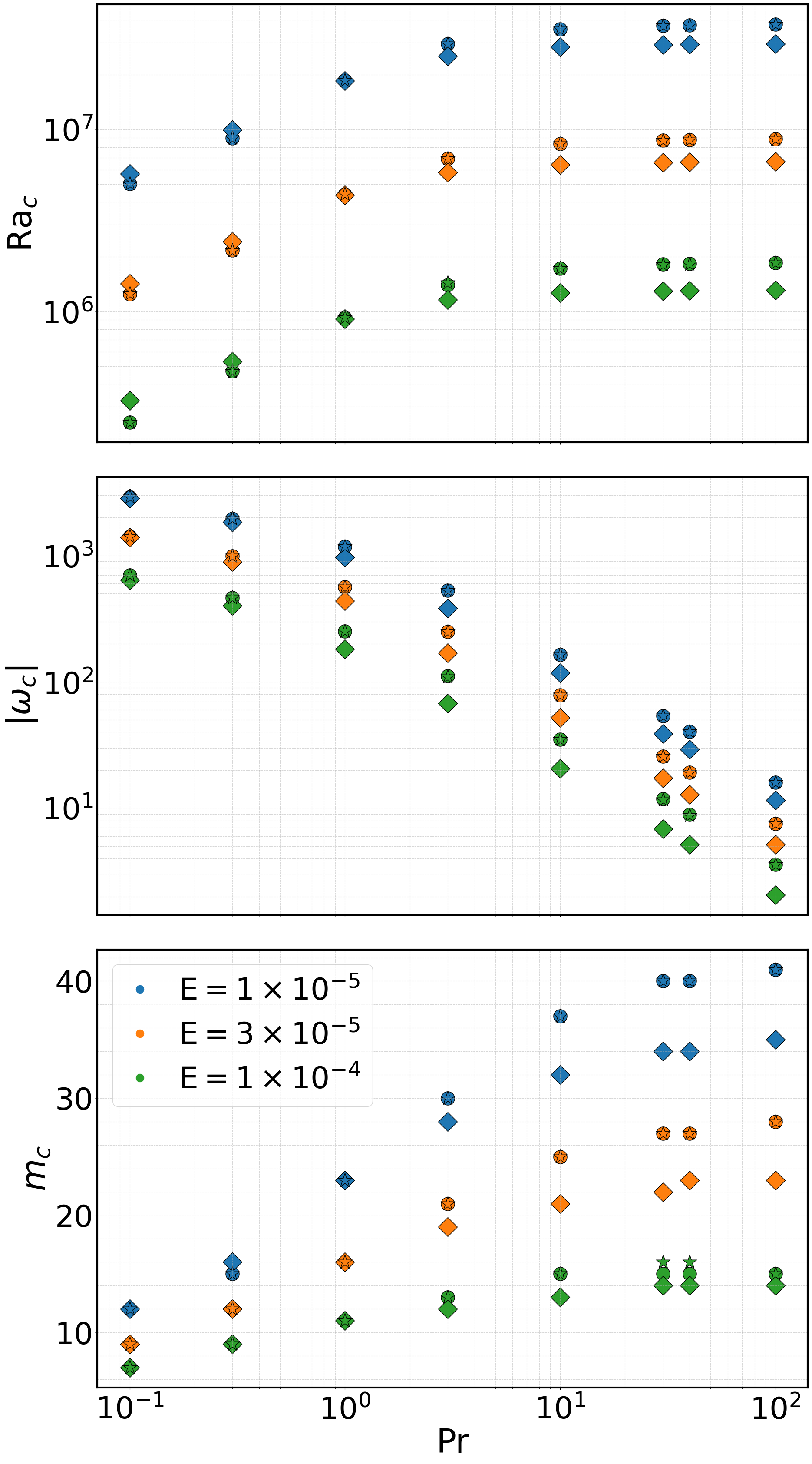}\label{IBCs}}
\caption{  \rs{\hl{Comparison} 
 of critical parameter values as functions of the Prandtl number, $\text{Pr}$, for several values of the Ekman number, E, and
different boundary conditions with radius ratio, $\chi = 0.35$. Results are shown for
each heating configuration: (\textbf{a}) differential heating \hl{(hollow symbols)}; (\textbf{b}) internal
heating \hl{(solid symbols)}. Boundary conditions are indicated by symbols: purely no‑slip ($\Diamond$); purely stress‑free ($\medcircle$); 
no-slip on the inner boundary and stress‑free on the outer boundary
($\largestar$). 
}}\label{fig4}
\end{figure}

{\emph{\hl{Comparison of critical values (stress-free boundary conditions). 
}}} 
Consistent with previous work \cite{dormy2004}, we find that purely stress-free boundary conditions lead to a larger $m_c$ and a larger $|\omega_c|$ for $\text{Pr}=1$, compared with purely no-slip conditions. We now confirm this result across the range of Prandtl numbers tested for both differential heating (Figure~\ref{fig4}a) and internal heating (Figure~\ref{fig4}b). In other words, for given E, Pr, and heating type (and $\chi=0.35$), we typically find $m_c^\text{NS}<m_c^\text{SF}$ and $|\omega_c|^\text{NS}<|\omega_c|^\text{SF}$. In fact, this is also true for the range of $\chi$ tested, and we shall return to this in Section \ref{sec:chires}.
For differential heating,
we reproduce the result found in \cite{dormy2004} and confirm that, for $\text{Pr}=1$, the critical Rayleigh number satisfies $\text{Ra}^\text{NS}_c>\text{Ra}^\text{SF}_c$. However, we note that in the internally heated case it has been observed that $\text{Ra}^\text{NS}_c<\text{Ra}^\text{SF}_c$ for $\text{Pr}=1$ \cite{dormy2004} and for larger Pr \cite{zhang1993}. This raises the question of whether or nor such a situation is possible in the differentially heated case and, if so, how it depends on Pr (and we shall return to this point later).

{\emph{\hl{Comparison of critical values (mixed boundary conditions). 
}}} We observe that the critical values for the mixed boundary-condition cases are very similar to those of the purely stress-free case (Figure \ref{fig4}a). This similarity arises as the convection columns dominantly interact with the outer boundary at the top and bottom of the column. Although in the differentially heated case the convection columns are found adjacent to the inner boundary, they are not wall-attached modes, and thus the inner boundary plays a weaker role. Consequently, the critical values of a mixed boundary case are closely approximated by the non-mixed boundary condition case with matching outer wall boundary condition (i.e.,~purely stress-free conditions for our choice of mixed conditions). Deviations in the mixed case (from the purely stress-free case) will depend on the relative strength of the no-slip inner boundary, which, in turn, depends on heating type, $\text{Pr}$, $\text{E}$, and $\chi$.

\begin{figure}[H]
\subfloat[\centering]{\includegraphics[width=0.48\linewidth]{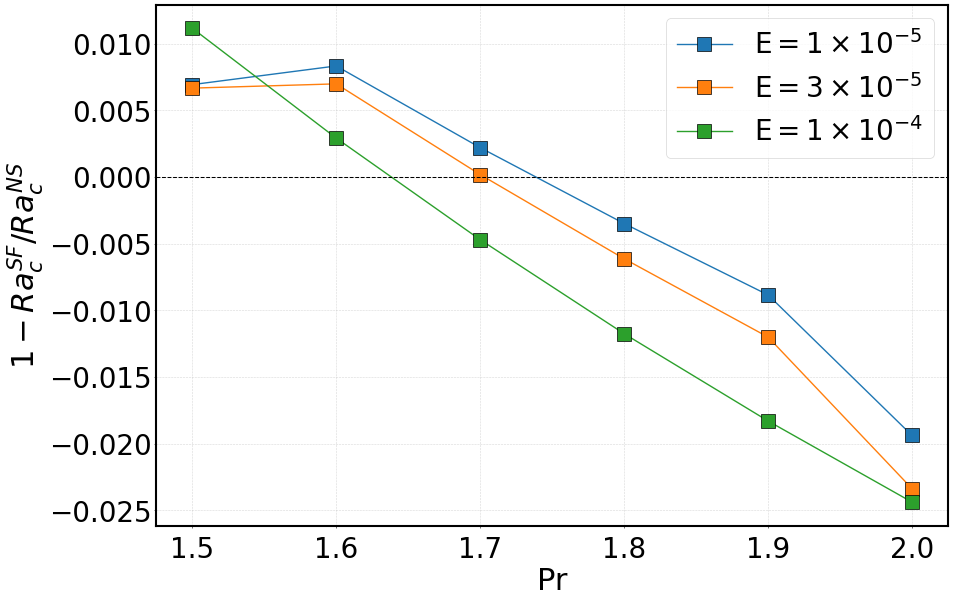}\label{BCsDiff_Pr}}
\subfloat[\centering]{\includegraphics[width=0.48\linewidth]{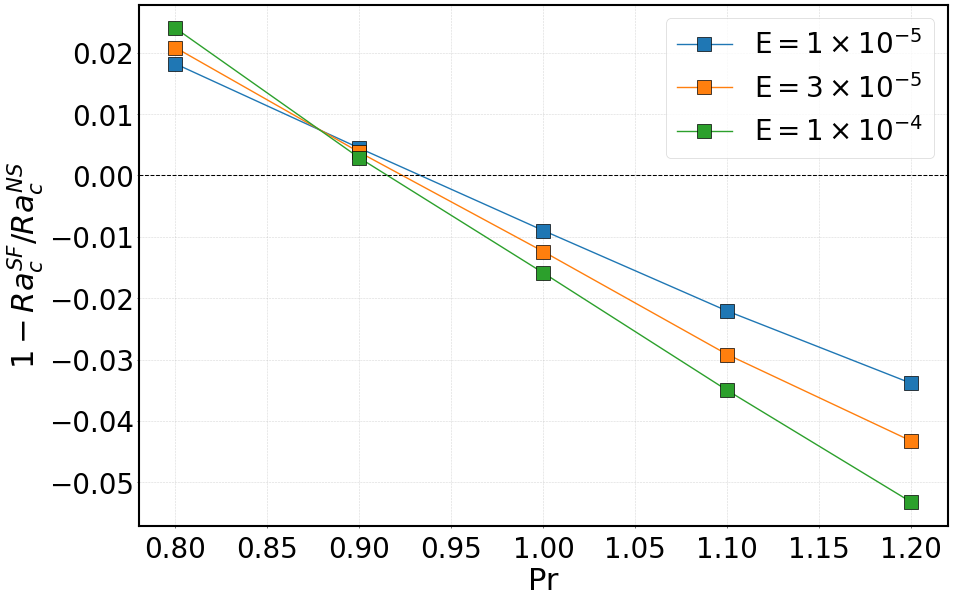}\label{BCsInt_Pr}}
\caption{
\hl{The} 
 relative difference between the critical Rayleigh numbers of purely no-slip and purely stress-free boundary conditions. The quantity $1-\text{Ra}^\text{SF}_c/\text{Ra}^\text{NS}_c$ is plotted against the Prandtl number, $\text{Pr}$, for different values of the Ekman number, $\text{E}$, with $\chi=0.35$ and the case of (\textbf{a}) differential heating and (\textbf{b}) internal heating. \label{BCs_Pr}}
\end{figure}

{\emph{\hl{Preference of boundary condition for onset. 
}}} We now examine, in more detail, how the relationship between  $\text{Ra}^\text{NS}_c$ and $\text{Ra}^\text{SF}_c$ varies with the Prandtl number for differential heating (and will consider the same effect for internal heating in Section \ref{sec:Prheating}). For $\text{Pr}<1$, we find that $\text{Ra}^\text{NS}_c>\text{Ra}^\text{SF}_c$ persists, with the difference between the two values increasing across the explored parameter space (see Figure \ref{fig4}a at small Pr). Hence, for low enough Pr, the Ekman boundary layer restrains the fluid, limiting the growth of instabilities, leading to a larger $\text{Ra}_c$. However, increasing the Prandtl number leads to the Ekman boundary layer having a destabilising effect on the flow, thereby reducing $\text{Ra}_c$ (see Figure \ref{fig4}a at large Pr).  
This effect occurs for the range of values of Ekman number tested, provided $\text{Pr}\gtrsim 1.7$. However, the precise transitional point in $\text{Pr}$-space depends slightly on the Ekman number (Figure \ref{BCs_Pr}a). We find that for larger values of $\text{E}$, the boundary layer remains sufficiently stabilising at increased Pr, meaning $\text{Ra}^\text{NS}_c>\text{Ra}^\text{SF}_c$ holds for larger values of $\text{Pr}$.
This transition does not affect the critical azimuthal wavenumber, $m_c$, or the critical frequency, $\omega_c$, as discussed earlier.
We can also highlight the decreasing influence of the Ekman boundary layer at lower Ekman number (since its thickness scales as $\text{E}^{1/2}$ \cite{cushman2011}). From Figure \ref{BCs_Pr}, we see that as $\text{E} \to 0$, ${\text{Ra}^\text{NS}_c}/{\text{Ra}^\text{SF}_c}\to 1$, ${m^\text{NS}_c}/{m^\text{SF}_c}\to 1$, and ${\omega^\text{NS}_c}/{\omega^\text{SF}_c}\to 1$. This can be most readily observed in Figure~\ref{BCs_Pr}b, where smaller values of $\text{E}$ tend closer to the value of $1-\text{Ra}^\text{SF}_c/\text{Ra}^\text{NS}_c=0$ across Pr-space.

{\emph{\hl{Flow patterns. 
}}} We find very little discernible difference in the pattern of the convection columns for different sets of boundary conditions. Therefore, we do not display further plots in this section beyond those shown in Figure~\ref{Mode Plots DNSNS}, which are for no-slip boundary conditions. The only key difference in the flow patterns are those which are expected from the critical values: the number of convection columns present in a given plot may be slightly different since $m_c$ varies (weakly) with selection of boundary condition.

\subsubsection{Effect of Heating Type}
\label{sec:Prheating}

 Previous studies have investigated the effect of different heating cases on critical values. The work of \cite{dormy2004} directly compared purely internal and purely differential for $\text{Pr}=1$. Subsequent work has focused on differentially heated systems, mainly within the parameter range of $\text{Pr}\in[0.01,10]$ \cite{fan2024, christensen2006}, whereas \cite{Silva2019} has looked at the internal heating case across a much larger range, $\text{Pr} \in [10^{-5},10^3]$. Here, we directly compare internal and differential heating while varying the Prandtl number, comparing how each heating case affects the critical values and interacts with boundary condition effects.

{\emph{\hl{Comparison of critical values. 
}}} 
\rt{Comparing Figure~\ref{fig4}a,b, it is clear that the trends with Pr are very similar for each heating type, although the magnitudes of the critical values can differ significantly. This is the case for all boundary conditions we have investigated and, as with differential heating, the mixed boundary conditions under internal heating provide similar results to the purely stress-free conditions. The critical Rayleigh numbers associated with internal heating are up to an order of magnitude larger than the equivalent in a differentially heated system. Conversely, the magnitudes of the critical frequencies are similar in the two heating cases, albeit slightly larger under internal heating. The dependence of the critical wavenumber on heating type is more complicated. For small Pr, values of $m_c$ are similar between the two cases, but for large Pr they become increasingly different as the number of modes increases significantly faster in the internally heated case. We shall return to this point below since it is related to a key difference between the two heating cases that we shall discuss next: the location of the onset of the instability.
}

\begin{figure}[H]
\includegraphics[width=0.68\linewidth]{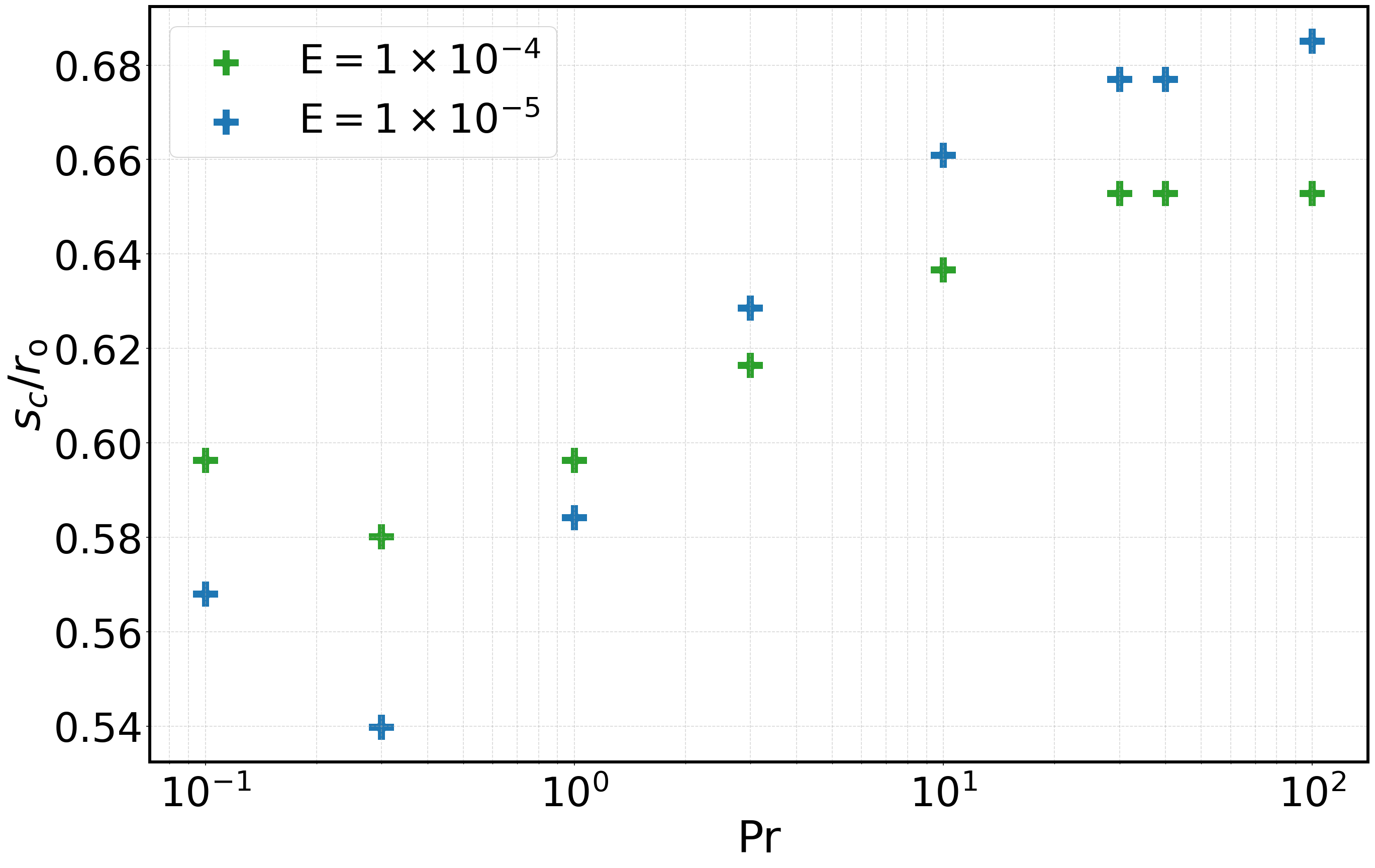}
\caption{\hl{The} 
 critical radius, $s_c$, as a function of the Prandtl number, $\text{Pr}$, for two values of the Ekman number, $\text{E}$, with $\chi=0.35$, purely no-slip boundary conditions, and for the case of internal heating.
\label{CritCylinderPr}}
\end{figure}

{\emph{\hl{Critical radius. 
}}} The (relative) resistance to instability in the internal heating case arises due to the outer boundary having the steepest temperature gradient (see Equation (\ref{nablaT0beta})). Therefore, the buoyancy forces are greatest at the outer boundary but must compete with a greater rotational constraint resulting from the increased slope \cite{busse1977}. Equivalently, the Taylor columns are disrupted as, near the outer wall, the fluid is forced to become more $z$-dependent. As a result, in the internal heating case the instability onsets away from the tangent cylinder at some critical radius, $s_c$, where the competition between buoyancy forces and rotational constraints is most favourable. Conversely, under differential heating, where the inner wall has the steepest temperature gradient, the instability always onsets adjacent to the tangent cylinder. This property is directly observed by comparing the upper half (for differential heating) and lower half (for internal heating) of the individual plots of \mbox{Figure \ref{Mode Plots DNSNS}}.
A natural question is how the critical radius varies with the control parameters of the system. The asymptotic value of the `global theory' of  \cite{dormy2004} in the case of $\text{Pr}=1$ was found to be $s_M/r_\text{o}\approx0.5915$, where it was also speculated (but not explored) that $s_c$ is dependent on the Prandtl number. In Figure \ref{Mode Plots DNSNS} (bottom half of each plot), it is notable that as $\text{Pr}$ is increased, the critical radius, $s_c$, becomes (marginally) further from the tangent circle, $r_\text{i}$.
Figure \ref{CritCylinderPr} demonstrates this behaviour over the full parameter range of Pr tested, where we find $0.54<s_c/r_\text{o}<0.69$ for the cases tested. There are several noteworthy trends to be highlighted. First, broadly speaking, increasing $\text{Pr}$ leads to the instability migrating further from the inner boundary. This is likely a result of the weakening effect of the Ekman boundary layer at larger Pr, allowing buoyancy forces to prevail over rotational constraints at a slightly larger $s_c$. Onset at a larger radius also leads to a larger number of modes since the circumference of the circle defined by $r=s_c/r_\text{o}$ increases; this explains why $m_c$ increases more rapidly with Pr in the internally heated case, as mentioned earlier.
Second, bucking the general trend seen in Figure \ref{CritCylinderPr}, for small enough Pr, the critical radius begins to increase. Hence, there exists a minimum value of $s_c$ as a function of Pr, and we find that this occurs when $\text{Pr}\approx0.3$. Third, the dependence on E demonstrates that larger rotation rates amplify the trends seen at both small and large Pr. This is likely because, as rotation is increased, modes become smaller and more localised to their inner/outer edge for small/large Pr (Figure \ref{Mode Plots DNSNS}).

{\emph{\hl{Preference of boundary condition for onset. 
}}} The behaviour described above persists across all boundary conditions, and similar trends in Pr are observed in the internally heated case as were found for the differentially heated case (see Figure~\ref{BCs_Pr}b). A notable difference, however, is the shift in transition where the no-slip case onsets at a lower $\text{Ra}_c$. In agreement with previous work \cite{zhang1993}, we find that, for internal heating, this transition occurs near $\text{Pr}\approx1$ (Figure~\ref{BCs_Pr}b), whereas for differential heating the equivalent value was $\text{Pr}\approx1.7$ (Figure~\ref{BCs_Pr}a). As with the differential heating case, we have found additional results in this region to more accurately determine where this transition occurs as a function of $\text{Pr}$ (and $\text{E}$).

{\emph{\hl{Flow patterns. 
}}} We have discussed some key effects of the choice of heating type on the form of the convection columns. Aside from the radial location of the instability and the change in the number of azimuthal modes, there is one further observation to be made from Figure \ref{Mode Plots DNSNS}. The convection cells are more radially elongated in the internally heated case, presumably because their extent is not inhibited by the inner boundary.

\subsection{Dependence on Radius Ratio}
\label{sec:chires}

All results presented in this subsection are for a fixed Prandtl number of Pr~=~1, since this allows for straightforward comparison with previous work.
As previous work \mbox{shows~\citep{dormy2004,Al-shamali2004, barik2024}}, varying $\chi$ has an effect on the critical values for the onset of convection. In the sections that follow, we present results for $\chi \in \{0.2,0.25,\dots,0.75,0.8\}$ for a variety of boundary conditions and heating types. However, we begin with a brief discussion of the case with no-slip boundary conditions and differential heating since this allows us to readily verify our results against previous studies.

\begin{figure}[H]
\centering
\subfloat[\centering\label{DNSNS Ra chi}]{\includegraphics[width=0.32\linewidth]{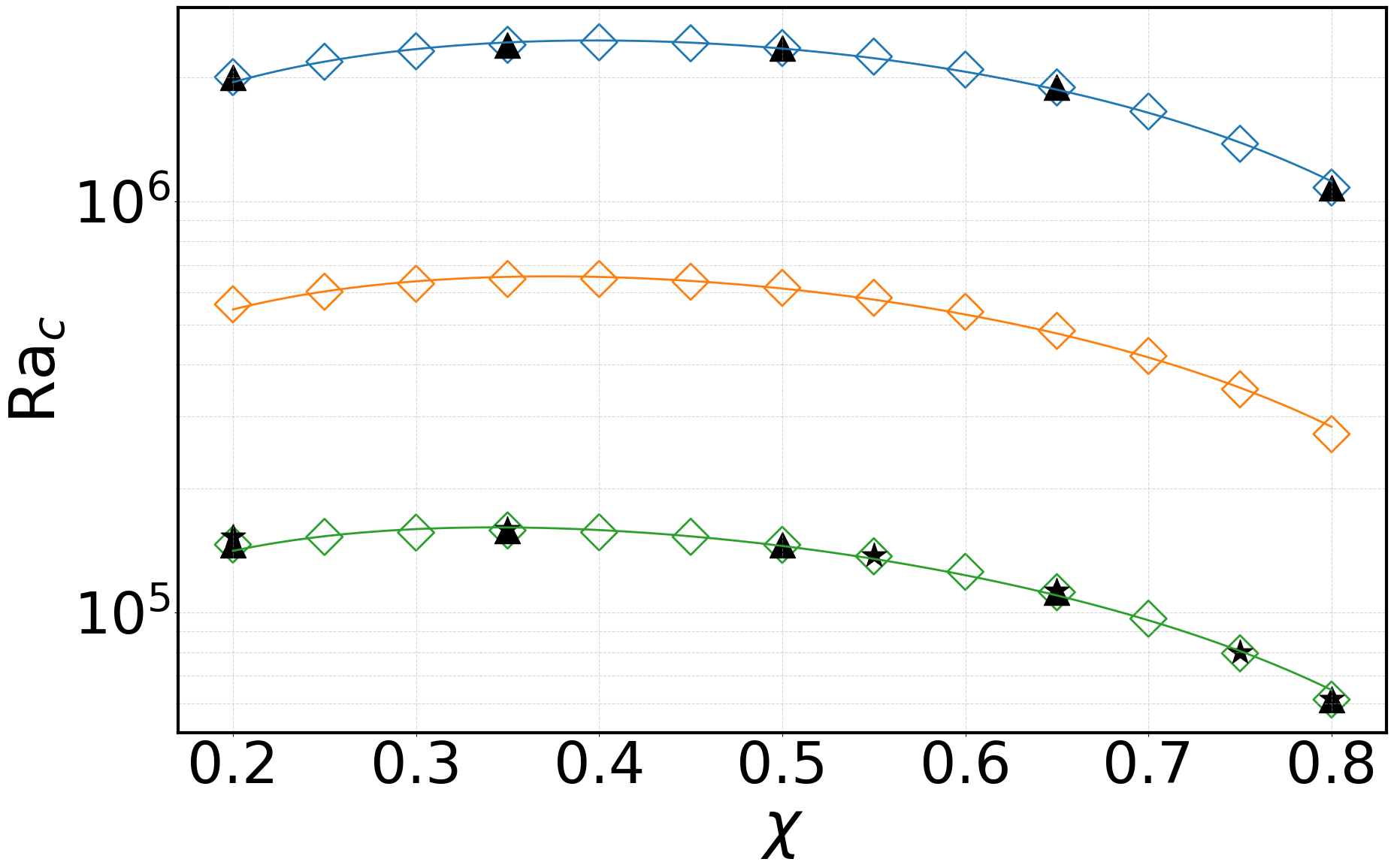}}
\subfloat[\centering\label{DNSNS w chi}]{\includegraphics[width=0.32\linewidth]{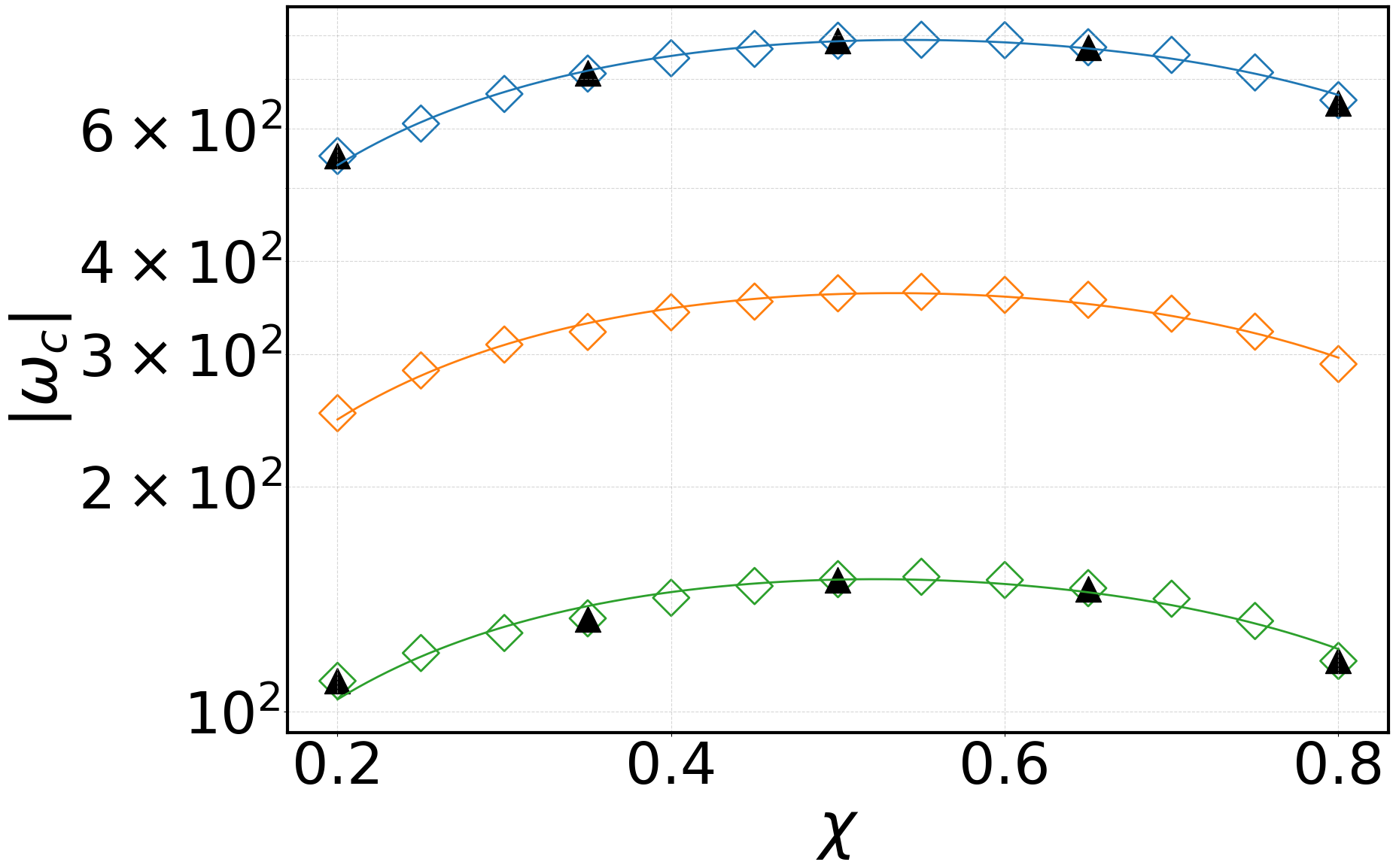}}
\subfloat[\centering\label{DNSNS m chi}]{\includegraphics[width=0.32\linewidth]{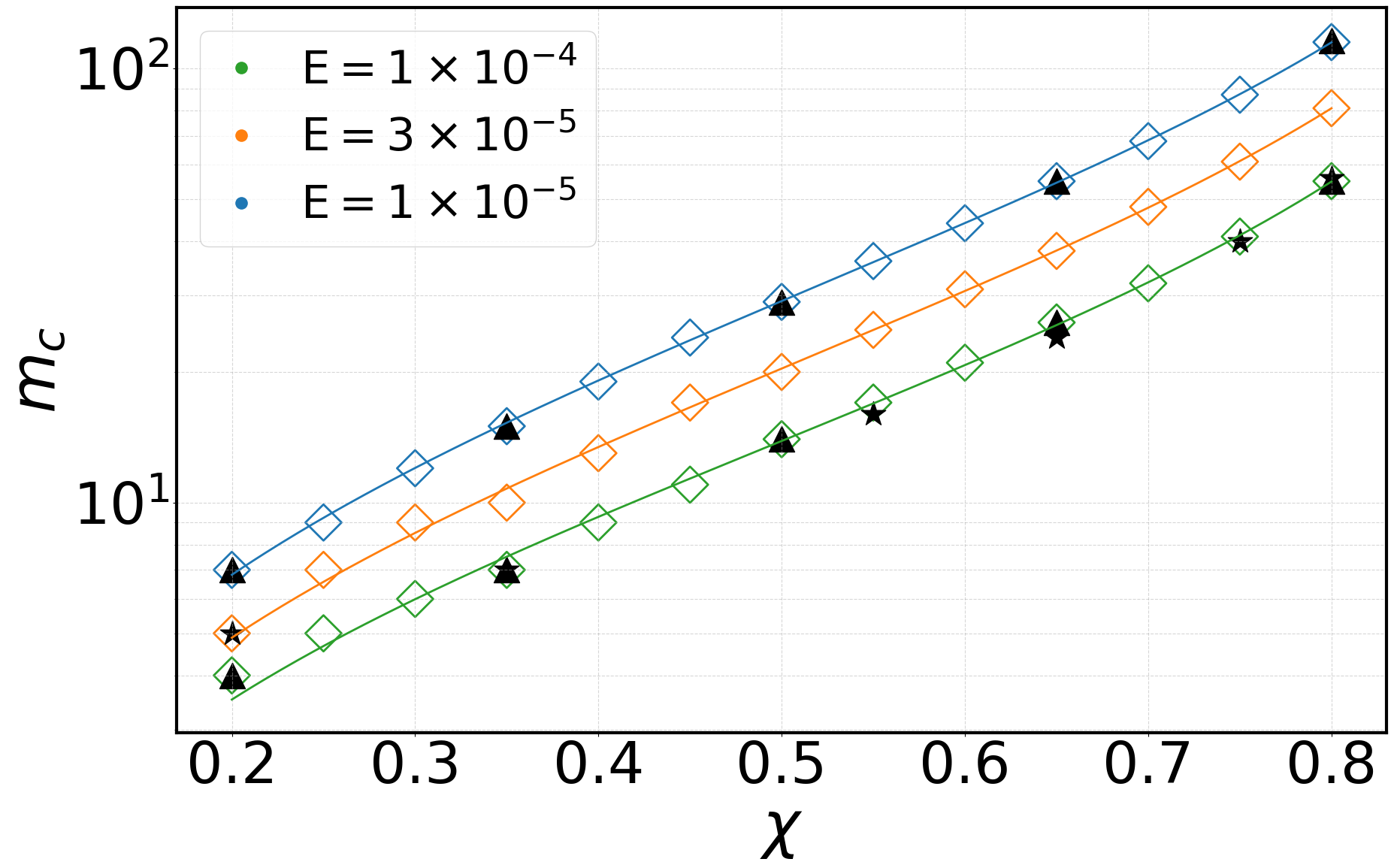}}
\caption{
Dependence of critical values on the radius ratio, $\chi$, for several values of the Ekman number, $\text{E}$, for the case of differential heating with purely no‑slip boundary conditions and Prandtl number, $\text{Pr}=1$.
(\textbf{a}) Critical Rayleigh number, (\textbf{b}) critical
  frequency, (\textbf{c}) critical azimuthal wavenumber.
  Critical values found by 
  Al-Shamali et. al. \cite{Al-shamali2004} ($\bigstar$) and Barik
  et. al. \cite{barik2024} ($\blacktriangle$)
  are also shown.
    The best-fit curves are given by Equation (\ref{critVal chi scaling}) with fitting parameters given in Table \ref{chi fitting exponents}.
}
\label{DNSNS M1}
\end{figure}

{\emph{\hl{Critical values. 
}}} Figure \ref{DNSNS M1} presents critical parameter values as a function of radius ratio found as part of this study.
As expected, based on the known scaling laws, all three critical parameters increase with E for all values of $\chi$.
Figure~\ref{DNSNS M1}a shows a small increase in the critical Rayleigh number until $\chi\approx0.4$ (though this value is weakly dependent on E), where it then starts to decrease.
\rt{This suggests that a spherical shell with a moderate aspect ratio ($\chi\approx0.4$) is least favourable to the onset of convection.
However, the trend seen in Ra$_c$ with $\chi$ is highly dependent on the definition of the Rayleigh number used for analysis (as alluded to in Section \ref{sec:altform}). We discuss this point further in Section \ref{ref:formulation}.}
 Figure~\ref{DNSNS M1}b shows that $|\omega_c|$ peaks at $\chi\approx0.55$. \rt{Given the instability onsets at $s\approx r_\text{i}$ (for differential heating), this suggests there is a favourable location in $s$-space for maximising the drift of the convection columns.}
Figure~\ref{DNSNS M1}c shows that the critical wavenumber increases with increasing $\chi$. This is partly due to the fact that, for differential heating, convection onsets in the form of columns at the tangent cylinder to the inner core, and as this inner core gets larger, the convection columns either need to stretch around it or split up into more columns, with the latter being most efficient.

{\emph{\hl{Scaling laws. 
}}} To plot the curves of best-fit for varying $\chi$, we again make use of scaling laws found in \cite{simitev2003}, 
which, for fixed E and Pr, provide:
\begin{flalign}
\label{critVal chi scaling}  C_\text{i}=a_\text{i}\bigg[\tan\bigg(\arcsin\bigg(\frac{1+\chi}{2}\bigg)\bigg)\bigg]^{b_\text{i}}\bigg(\frac{1+\chi}{1-\chi}\bigg)^{c_\text{i}},
\end{flalign}
where $C_1=\chi^{-1}\text{Ra}_c$, $C_2=m_c$, and $C_3=\omega_c$. The fitting parameters $a_\text{i}$, $b_\text{i}$, and $c_\text{i}$ for Figure~\ref{DNSNS M1} are given in Table \ref{chi fitting exponents}. Note, there are similarities between the expression for $C_1$ in (\ref{critVal chi scaling}) and the relationship given by Equation (9) in \cite{Al-shamali2004}, especially in the form of the part with the $c_1$ exponent. However, we find a more suitable fit for critical values when using the above scaling law.
This 
may arise due to differences in the range of $\chi$ values studied.
We keep to intermediate values of $\chi$ where the law from \cite{simitev2003} (based on the local annulus model for convection outside the tangent cylinder) is appropriate. On the other hand, ref.~\cite{Al-shamali2004} also test extreme values of $\chi$ appropriate to 
the thickest and thinnest shell cases where the changes  in $\text{Ra}_c$ are more dramatic and require special treatment by their revised scaling law. 
We do not find each fitted parameter, $a_i$, $b_i$, and $c_i$, tending towards the proposed values provided by  \cite{simitev2003}; however, we should not necessarily expect to find agreement since those values are derived for internal heating and are specifically finding the scaling for small Ekman numbers. Conversely, here, we use differential heating and retain finite, moderate values of E. Lower values of the Ekman number and more extreme values of $\chi$ would be needed to further explore these scaling laws.
Nevertheless, for each fixed value of $\chi$, we do recover the expected scalings, $\text{Ra}_c\sim \text{E}^{-4/3}$, $m_c\sim \text{E}^{-1/3}$ and $\omega_c\sim \text{E}^{-2/3}$, as $\text{E}\to0$.

\begin{table}[H]
  \caption{Fitting parameters $a_i$, $b_i$, and $c_i$ for the critical value best-fit curves from Figure \ref{DNSNS M1} (for a system with $\text{Pr}=1$, differential heating, and purely no-slip boundaries). The fitting parameters are used in Equation (\ref{critVal chi scaling}), where $C_1=\text{Ra}_c$, $C_2=m_c$, and $C_3=\omega_c$.
    }
    \label{chi fitting exponents}
  \hspace*{-1.5cm}
  \begin{tabularx}{\fulllength}{CcCCccCcCC}
  \toprule 
        \textbf{$\text{E}$} & \boldmath{$a_1$} & \boldmath{$b_1$} & \boldmath{$c_1$}
        & \boldmath{$a_2$} & \boldmath{$b_2$} & \boldmath{$c_2$}
        & \boldmath{$a_3$} & \boldmath{$b_3$} & \boldmath{$c_3$} \\
        \midrule
        $10^{-4}$  &  $3.407\times10^{5} $ &$-2.355$ & $0.123$
        & $4.172\times10^{2} $ &$10.147 $ & $-4.273 $
        & $9.751\times10^{4} $ &$13.165 $ & $-7.402$ \\
        \midrule
        $3\times10^{-5}$ &$3.540\times10^{6} $ &$-0.390$ & $-0.920 $
        & $1.235\times10^{3} $ &$11.636 $ & $-5.081 $
        & $2.477\times10^{5} $ &$13.320 $ & $-7.470$ \\
        \midrule
        $10^{-5}$  &  $2.662\times10^{7} $ &$1.096$ & $-1.703 $ 
        & $2.665\times10^{3} $ &$12.486$ & $-5.553$
        & $3.746\times10^{5} $ &$12.620 $ & $-7.059$ \\
        \bottomrule
    \end{tabularx}
  \end{table}

{\emph{\hl{Comparison with previous studies. 
}}
Figure \ref{DNSNS M1} also shows that, as was the case when varying Pr, the critical values we have calculated as a function of $\chi$ agree very well with previous studies (after suitable conversions; see Section \ref{sec:altform}). Both Ra$_c$ and $|\omega_c|$ show excellent agreement with the results of \cite{Al-shamali2004,barik2024}, and the discrete parameter, $m_c$, is mostly in agreement, especially with the more recent study \cite{barik2024}.
}

\begin{figure}[ht!]
    \centering
    \subfloat[\centering\label{DNSNS v 2}]
    {\includegraphics[width=0.24\linewidth]{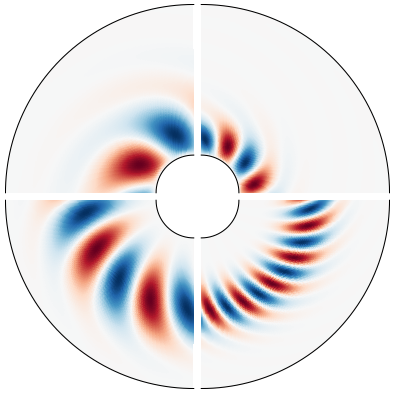}}
    \subfloat[\centering\label{DNSNS v 45}]
    {\includegraphics[width=0.24\linewidth]{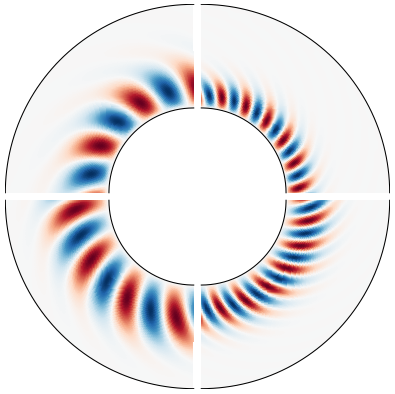}}
    \subfloat[\centering\label{DNSNS v 65}]
    {\includegraphics[width=0.24\linewidth]{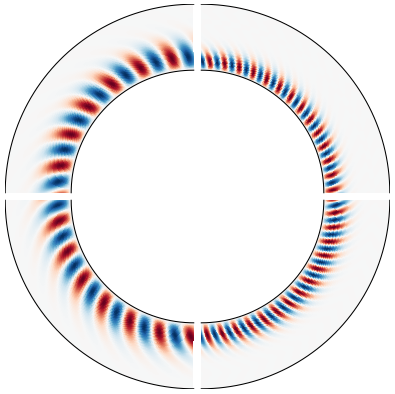}}
    \subfloat[\centering\label{DNSNS v 8}]
    {\includegraphics[width=0.24\linewidth]{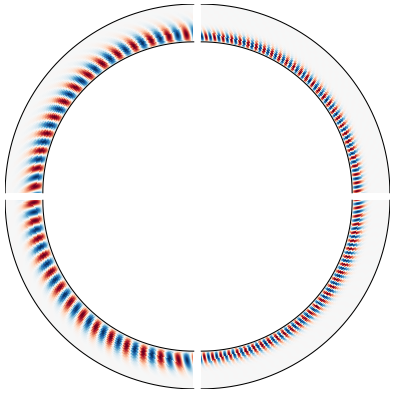}}
    \caption{\hl{The} 
 radial component of the velocity field at onset plotted in an equatorial slice for purely no-slip boundary conditions, $\text{Pr}=1$,
    differential heating (top half), internal heating (bottom half), $\text{E}=10^{-4}$ (left half of each panel), $\text{E}=10^{-5}$ (right half of each panel), and for (\textbf{a}) $\chi=0.2$, (\textbf{b}) $\chi=0.45$, ({\textbf c})~$\chi=0.65$, and (\textbf{d}) $\chi=0.8$. \hl{Red and blue represent positive and negative values, respectively.}}
    \label{vel plots DNSNS}
\end{figure}

{\emph{\hl{Flow patterns. 
}}} The equatorial sections displayed in Figure~\ref{vel plots DNSNS} show the form of the convection columns.
\rt{In the differentially heated case (top half of each plot), the modes are adjacent to the inner boundary;
comparison with the internally heated case will be made in Section \ref{sec:chiheating}.
The plots highlight the dependence of the size of the flow structures on both Ekman number (compare left and right of each plot) and radius radio (shown in individual plots).
As $\chi$ is increased, the number of columns increases (even if E and Pr are held constant), with our most extreme case (right part of Figure \ref{vel plots DNSNS}d) showing extremely small structures. This highlights the requirement for azimuthal spatial high resolution in nonlinear rotating convection and dynamo simulations.
}
\vspace{-12pt}

\subsubsection{Effect of Mechanical Boundary Conditions}
\label{sec:chiBCs}

\begin{figure}[H]
    \subfloat[\centering\label{fig:D BCs chi}]
    {\includegraphics[width=0.5\linewidth]{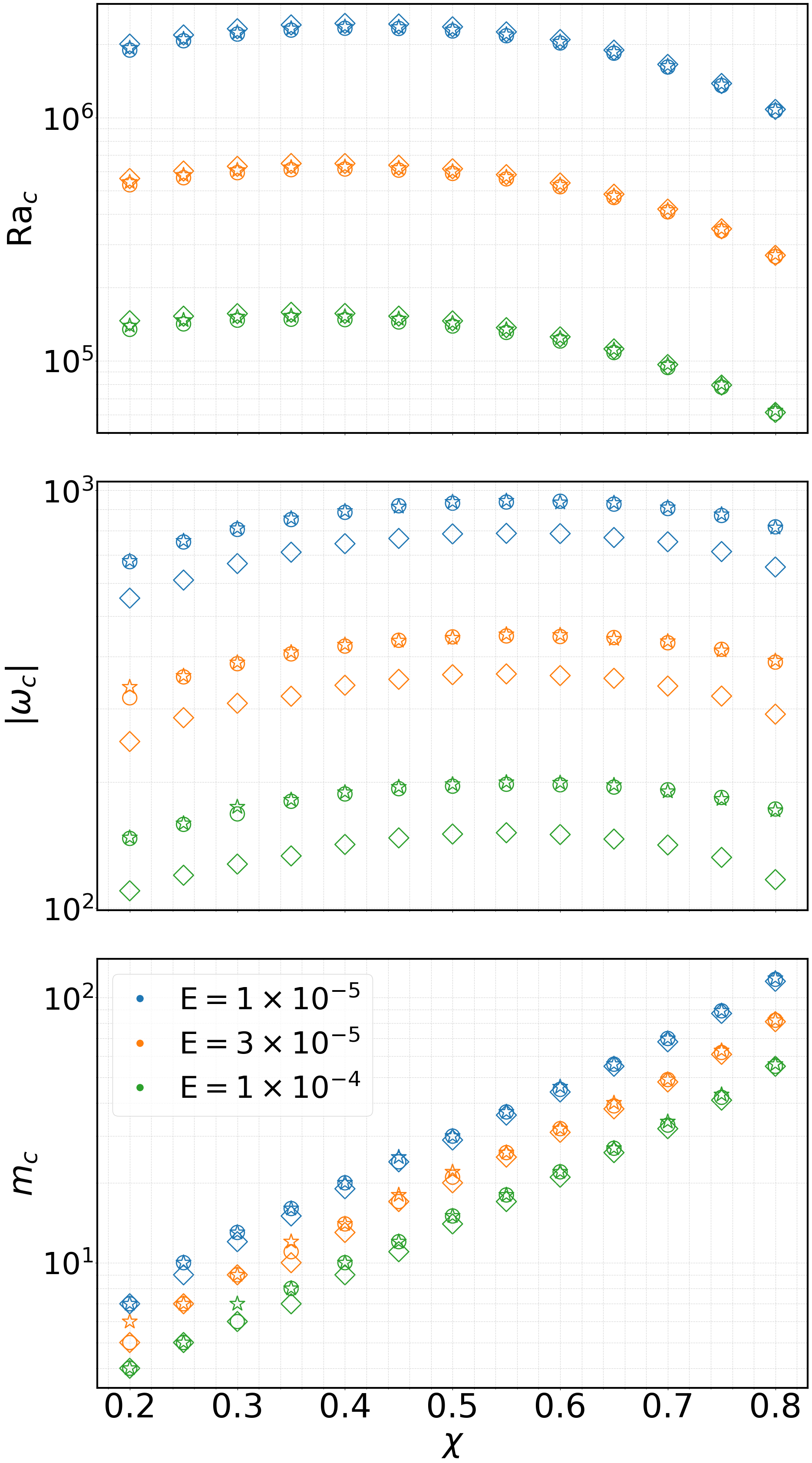}}
    \subfloat[\centering\label{fig:I BCs chi}]
    {\includegraphics[width=0.5\linewidth]{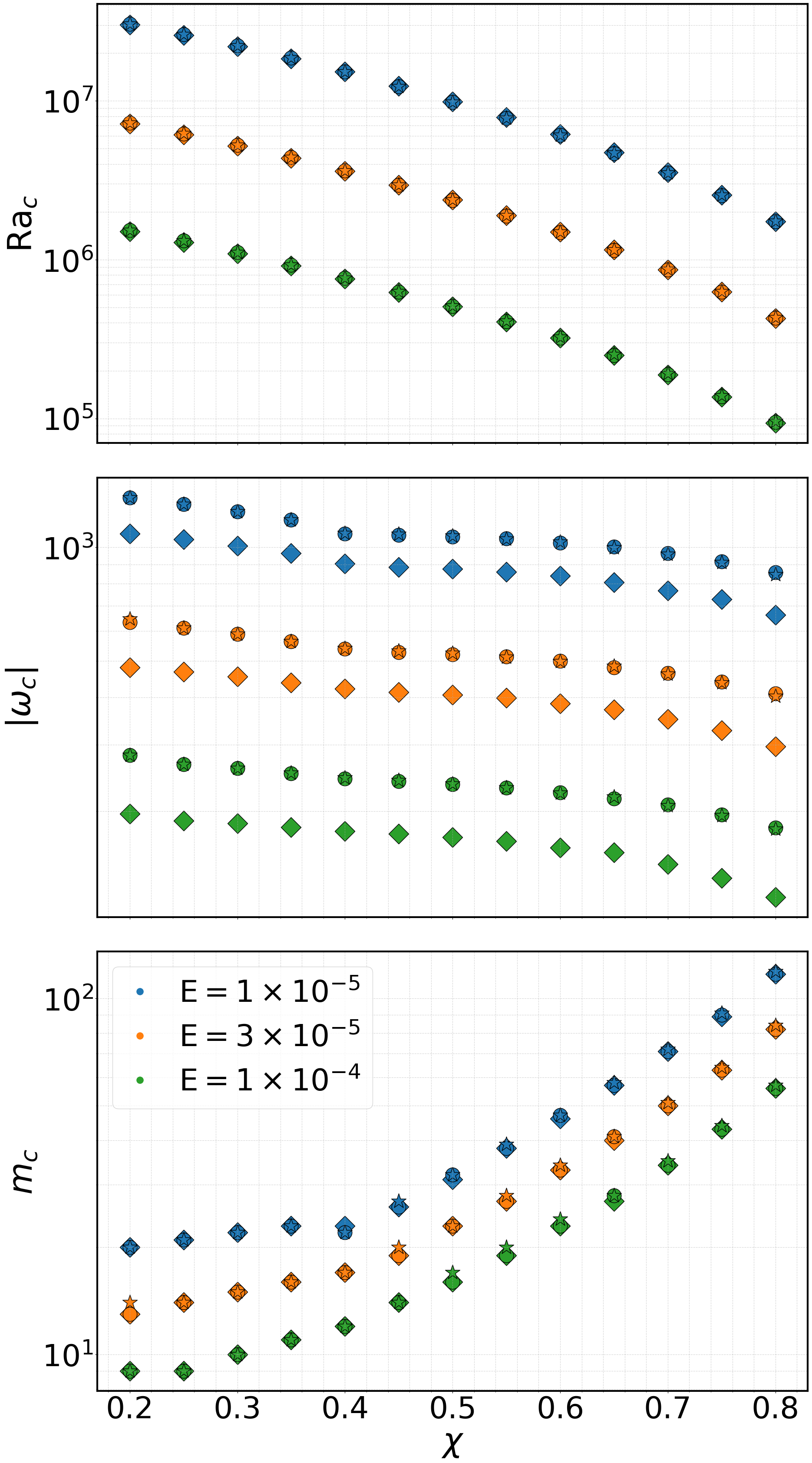}}
    \caption{\hl{Comparison} 
 of critical parameter values as functions of radius ratio, $\chi$, for several values of the Ekman number, $\text{E}$, and different boundary conditions with Prandtl number, $\text{Pr}=1$.
Results are shown for each heating configuration: (\textbf{a}) differential heating \hl{(hollow symbols)}; (\textbf{b}) internal heating \hl{(solid symbols)}. Boundary conditions are indicated by symbols: purely no‑slip ($\Diamond$); purely stress‑free ($\medcircle$); 
no-slip on the inner boundary and stress‑free on the outer boundary
($\largestar$).}
    \label{fig: NS SF and M w/ chi}
\end{figure}

{\emph{\hl{Comparison of critical values (stress-free boundary conditions). 
}}} 
Figure \ref{fig: NS SF and M w/ chi} compares the effect of various mechanical boundary conditions on the onset of convection, as a function of radius ratio. 
Figure \ref{fig: NS SF and M w/ chi}a shows that purely stress-free boundary conditions cause $\text{Ra}_c$ to decrease slightly, relative to the purely no-slip case.
Hence, with differential heating and for each $\chi$, convective onset is always easiest with stress-free conditions (at least for Pr~=~1).
This suggests that when using no-slip boundary conditions, the Ekman boundary layers have a stabilising effect on the fluid. This is in contrast to convective onset as a function of Pr, where the preference of boundary conditions varies (see Section \ref{sec:PrBCs}). 
The critical wavenumber (Figure \ref{fig: NS SF and M w/ chi}a) is altered slightly by the change in boundary conditions, with no-slip boundaries favouring a smaller number of modes (as seen earlier and in previous work). This remains the case across all values of $\chi$, for differential heating, although the difference becomes wider at large $\chi$. 
For stress-free boundary conditions, the magnitude of the critical frequency is larger than for no-slip boundary conditions (Figure \ref{fig: NS SF and M w/ chi}a). This is because the boundary layers that arise under no-slip boundary conditions inhibit the flow, causing a lower magnitude frequency.
\rt{In summary, for given E, $\chi$, and heating type (and Pr~=~1) we always find $m_c^\text{NS}<m_c^\text{SF}$ and $|\omega_c|^\text{NS}<|\omega_c|^\text{SF}$. Given the results of Section~\ref{sec:Pr}, it is likely that this remains the case across the full range of Pr and $\chi$ (i.e.,~for those combinations of Pr, $\chi$ is not included in our study).  
Similarly, for \hl{differential heating}, 
 for given E, $\chi$ (and Pr~=~1), we always find Ra$_c^\text{SF}<$Ra$_c^\text{NS}$. 
}

{\emph{\hl{Comparison of critical values (mixed boundary conditions). 
}}}
Figure \ref{fig: NS SF and M w/ chi} shows that the critical values for the case of mixed boundary conditions are very similar to the purely stress-free set-up.
For a fixed choice of $\chi$, the value of $\text{Ra}_c$ sits between the two other values, and the critical wavenumber and frequency are each almost identical to their equivalent values in the purely stress-free case.
Thus, the condition on the outer boundary, where the instability has its primary contact with the boundaries, is predominantly controlling the properties of the instability.
Hence, the introduction of a no-slip \emph{inner} boundary has a minimal effect to the result of the purely stress-free case. This was also observed in Section \ref{sec:Pr} but is now confirmed here for a range of $\chi$.
As $\chi$ is increased, the three values of $\text{Ra}_c$ (found under the three different choices of boundary conditions with differential heating) each tend towards the same value, indicating the diminishing role of the boundary conditions at large $\chi$.

{\emph{\hl{Flow patterns. 
}}
We do not display further plots of the flow structures in this section beyond those shown in Figure \ref{vel plots DNSNS}, which are for no-slip boundary conditions. This is because, as was the case in Section \ref{sec:Pr}, the only observable  difference in the flow patterns between cases with different boundary conditions is the number of columns, as expected  
because $m_c$ varies (weakly) with selection of boundary condition.
}

\subsubsection{Effect of Heating Type}
\label{sec:chiheating}

We now consider the case of internal heating, whilst also retaining the option to vary the boundary conditions. This section complements the numerical study contained in~\cite{dormy2004}, which considered two specific values of $\chi$ when comparing differential and internal heating.

{\emph{\hl{Comparison of critical values. 
}}}
 Figure \ref{fig: NS SF and M w/ chi}b shows the critical Rayleigh number decreasing monotonically as $\chi$ is increased. This differs from the differentially heated case where the system had a peak in Ra$_c$ at $\chi\approx0.4$. The trend in the internally heated case suggests that increasing $\chi$ always aids the onset of convection.
 The general trend in critical wavenumber (Figure~\ref{fig: NS SF and M w/ chi}b) is unchanged from the differentially heated case, retaining the significant increase in $m_c$ as $\chi$ increases.
 \rt{Noteworthy, however, is the slight `kink' in the values at $\chi\approx0.4$, which can be attributed to the instability becoming detached from the inner boundary as $\chi$ is reduced. We shall return to this point below.
This `kink' can also be observed in the plot for the critical frequency (especially for lower E) and, unlike the differentially heated case, there is no peak in $|\omega_c|$ at $\chi\approx0.55$.}

\begin{figure}[H]
    \includegraphics[width=0.68\linewidth]{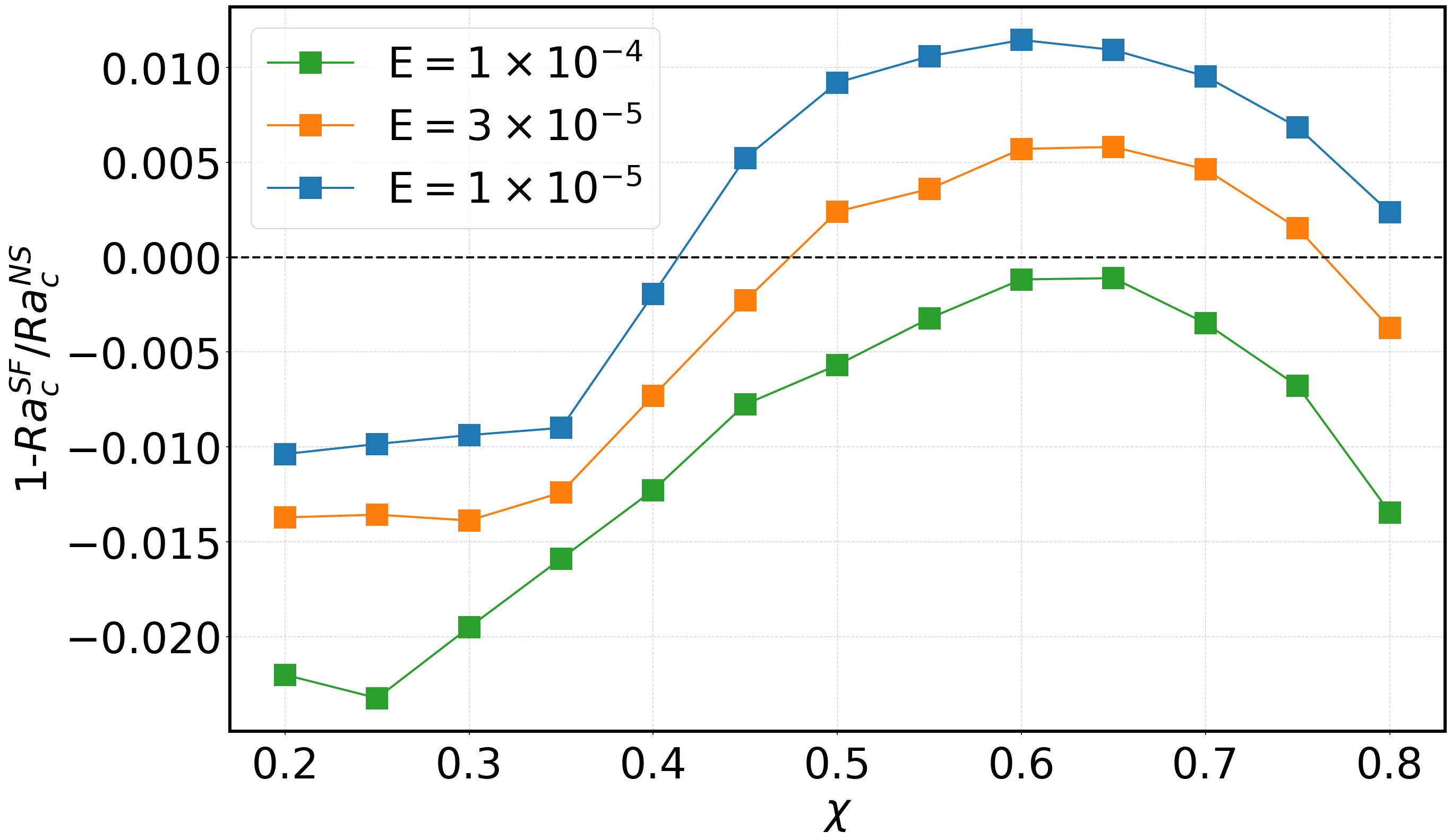}
    \caption{\hl{The} 
 relative difference between the critical Rayleigh numbers of purely no-slip and purely stress-free boundary conditions. The quantity $1-\text{Ra}_c^\text{SF}/\text{Ra}_c^\text{NS}$ is plotted against radius ratio $\chi$ for different values of the Ekman number $\text{E}$ with Prandtl number $\text{Pr}=1$ and the case of internal heating.
    }
    \label{fig:Transition with chi}
\end{figure}

{\emph{\hl{Preference of boundary condition for onset. 
}}}
For the internally heated system, the critical wavenumber and critical frequency are larger under purely stress-free or mixed boundary conditions compared to purely no-slip conditions. This is for the same reasons outlined previously in the case of differential heating.
On the other hand, as also found in the case of varying Pr (see Section \ref{sec:Pr}), there can be a transition in $\chi$-space where the preferred boundary condition to enable convection is reversed. In other words, the quantity \linebreak  $1-\text{Ra}_c^\text{SF}/\text{Ra}_c^\text{NS}$ may change sign as $\chi$ is varied, as shown in Figure \ref{fig:Transition with chi}.
\rt{Note, however, that as a function of $\chi$, this transition only occurs under internal heating. For differential heating we always find Ra$_c^\text{SF}<$Ra$_c^\text{NS}$.
The trend of $1-\text{Ra}_c^\text{SF}/\text{Ra}_c^\text{NS}$ in $\chi$ seen in Figure \ref{fig:Transition with chi} is similar for all values of E tested. However, this masks a subtlety since the number of times the quantity crosses zero is highly dependent on the Ekman number.}
For small values of $\chi$, no-slip boundary conditions give a lower critical Rayleigh number for all E. Then, for a sufficiently low value of E ($\le3\times10^{-5}$), there is a range of $\chi$ where stress-free boundary conditions become preferred. 
At large $\chi$, for $\text{E}=3\times10^{-5}$ only, the sign changes again so that no-slip boundary conditions are once again preferred. 
\rt{Note that for a wider range of values of radius ratio than studied here (i.e.,~for $\chi>0.8$), the final observation at large $\chi$ may also be true for lower values of E. The complex behaviour shown in Figure \ref{fig:Transition with chi} is likely also dependent on Pr and, as such, may warrant further study as a function of both E and Pr. Nevertheless, the trend we can draw out from Figure \ref{fig:Transition with chi} is that, at the low values of E relevant to planetary interiors, it is likely that no-slip (stress-free) boundaries are preferred at small (large) enough $\chi$.  
}

{\emph{\hl{Location of convection. 
}}}
As noted earlier, when using internal heating, it is possible for convection to onset away from the tangent cylinder. Owing to the strengthening of the temperature gradient with radius, the instability manifests at a critical cylinder of radius, $s_M\approx0.5915 r_\text{o}$, provided $\chi$ is small enough. The value of $s_M$ was asymptotically derived as part of a `global theory' \citep{jones2000,dormy2004}.
The critical cylinder matches that of the full sphere (where the differentially heated case cannot arise), since this position is unaffected when a sufficiently small inner core is introduced.
From Figure \ref{vel plots DNSNS}a (lower half for internal heating), it is immediately apparent, in the case of $\chi=0.2<s_M/r_\text{o}$, that the convection columns congregate at a radial value away from the inner core, as predicted. As $\chi$ is increased, the effect remains but becomes less obvious at $\chi=0.45$ (Figure \ref{vel plots DNSNS}b).
Naturally, as the inner core increases in size, it eventually encroaches on the columns found within the critical cylinder. \rt{Thus, for large enough $\chi$, and certainly for $\chi>s_M/r_\text{o}$, the optimal location for convection to manifest is at some critical radius, $s_L$, adjacent to the tangent cylinder (mirroring the situation for differential heating) \cite{dormy2004}.
The value of $s_L$ was found in the earlier `local theory'~\cite{roberts1968,busse1970,busse1977}.}
In Figure \ref{vel plots DNSNS}c,d (where $\chi>s_M/r_\text{o}$), the two heating cases appear indistinguishable since columns form adjacent to the tangent cylinder regardless.

\begin{figure}[H]
    \includegraphics[width=0.68\linewidth]{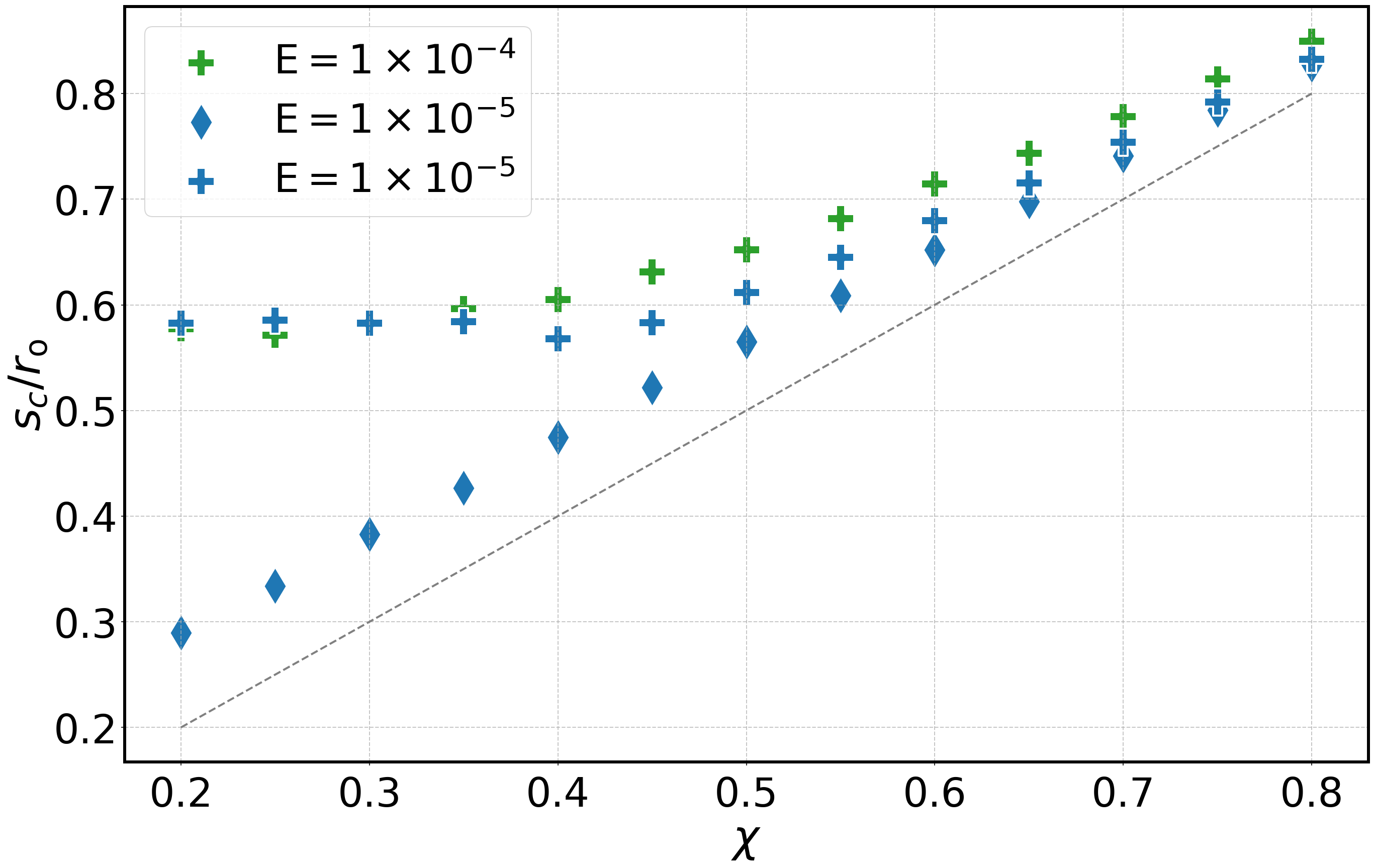}
    \caption{\hl{The} 
 critical radius, $s_c$, as a function of $\chi$ for two values of the Ekman number, $\text{E}$, with $\text{Pr}=1$, purely no-slip boundary conditions, and for the case of internal heating ($\mathbf{+}$) and differential heating ($\Diamond$).
    The grey dashed line is given by $s_c/r_\text{o}=\chi$, indicating the position of the inner boundary.
    }
    \label{fig:critical cylinder chi}
\end{figure}

{\emph{\hl{Critical radius. 
}}
In our work, the critical radius was discussed in Section \ref{sec:Prheating} for the fixed radius ratio of $\chi=0.35$. We now focus on the dependence of the critical radius with $\chi$.}
 Previous work \cite{dormy2004} studied only two values of $\chi$ and confirmed that $s_c\approx s_M$ for $\chi=0.35$ and $s_c\approx s_L$ for $\chi=0.65$.
  We track the critical radius across a range of values of $\chi$ for both internal heating and (the more trivial) differential heating, as shown in Figure~\ref{fig:critical cylinder chi}.
\rt{For the differentially heated system, the critical radius closely follows the curve $s/r_\text{o}=\chi$, as expected.}
Also notable is that, as $\chi$ increases, the critical cylinder moves closer to the tangent cylinder. \rt{This is a result of the columns reduction in radial (and azimuthal) extent as $\chi$ is increased (as seen clearly in Figure \ref{vel plots DNSNS}), allowing a closer proximity between the centre of the cells and the inner boundary. 
For the internally heated system, two different behaviours can be observed, as expected. For small $\chi$, the critical radius satisfies $s_c\approx s_M\approx0.59r_\text{o}$ for both values of E shown. For large $\chi$, we instead find that the critical radius moves outwards linearly with $\chi$. It is notable that the behaviour of the two heating types is different, even for $s>s_M$. Columns reside farther from the inner boundary in the internally heated case and the location of the critical cylinder in the two heating cases does not coincide (except perhaps at very large $\chi$). The smaller extent of the columns at low Ekman number also draws the columns closer to the tangent cylinder.
Lastly, the transition from $s_c\approx s_M$ to $s_c\approx s_L$ in the interval $0.4634<s/r_\text{o}<0.5004$ (where neither the `global' nor the `local' theory applies \cite{dormy2004}) appears to be smooth.
}

\subsection{Dependence on Nondimensionalisation}
\label{ref:formulation}

\begin{figure}[H]
    \includegraphics[width=0.9\linewidth]{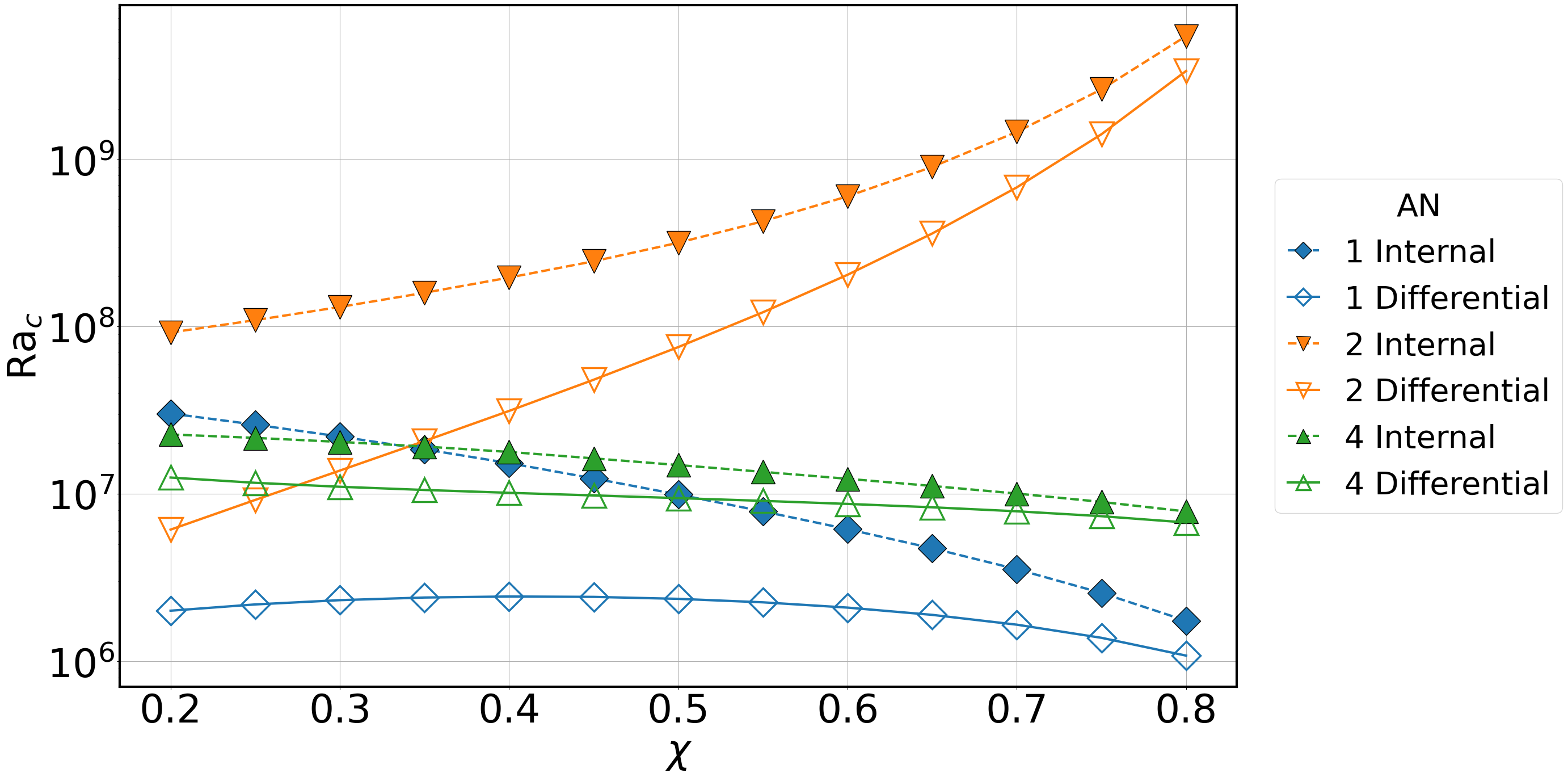}
    \caption{
    Dependence of the critical Rayleigh number on radius ratio for various definitions of Ra$_c$ based on nondimensional set-ups discussed in Section \ref{sec:altform}.
    All results shown are for $\text{E}=10^{-5}$ and $\text{Pr}=1$ with purely no-slip boundary conditions.
    Results are displayed for \textbf{AN-1} (as used in this work), \textbf{AN-2} (as used by \cite{dormy2004}), and \textbf{AN-4} (as used by \cite{Al-shamali2004,barik2024}). The curves are formed by connecting the points of each case.
    }
    \label{fig: compare models}
\end{figure}

As previously discussed, we confirmed the veracity of our results against previous studies \cite{dormy2004,Al-shamali2004,barik2024} using conversions given in Section \ref{sec:altform}, with selected points shown in Figures \ref{DNSNS PR} and \ref{DNSNS M1}.
\rt{As shown in Table \ref{Table: Conversions}, conversion of Ra$_c$ between alternative non-dimensionalisations requires different factors involving $\chi$.}
Hence, beyond simply confirming our results, it is also natural to convert our raw critical Rayleigh numbers to their equivalent values in the alternative nondimensionalisations and compare the trend in $\text{Ra}_c$ with $\chi$ in each.
Figure \ref{fig: compare models} shows the dependence of $\text{Ra}_c$ on radius ratio for each of the nondimensionalisations.

{\emph{\hl{Comparison with \textbf{AN-4}. 
}}}
\rt{First, we re-emphasise that, for \textbf{AN-1} (as used in our study) and under differential heating, Ra$_c$ decreases as $\chi$ increases but only after a shallow peak is reached at $\chi\approx0.4$.
Conversely, Ra$_c$ is a monotonically decreasing function of $\chi$ under internal heating.}
The general trend of Ra$_c$ decreasing with $\chi$ is also  
maintained for both heating cases for \textbf{AN-4} (used by \cite{Al-shamali2004,barik2024}).
\rt{The notable difference between the trends of {\bf AN-1} and {\bf AN-4} is the steepness of the curves for internal heating.}
{\bf AN-1} uses the temperature gradient (evaluated at $r_\text{o})$ within the definition of Ra (the key difference with {\bf AN-3} and {\bf AN-4}). Since the temperature gradient under internal heating scales with $r$, and hence becomes strongest at the outer boundary, Ra$_c$ is significantly reduced as the onset location moves towards the outer boundary. This effect is far more modest when the Rayleigh number is measured using the temperature difference because the geometric factor of $\Gamma(\chi)$ within the heat equation reduces the effect of buoyancy, leading to larger values of Ra$_c$.
Note that values of Ra$_c$ for nondimensionalisation \textbf{AN-3} 
 very closely follow those of \textbf{AN-4}.
\rt{Based on the trends in Ra$_c$ with $\chi$ shown by {\bf AN-1}, {\bf AN-3}, and {\bf AN-4}, a case can be made for thinner shells aiding the onset of convection. Indeed, ref.~\cite{Al-shamali2004} argue the effect is explained by the result that convection is easiest in cylinders of large aspect ratio \cite{zhanggreed1998}, owing to the reduced size of the boundary layers within the total volume of the fluid. Boundary layer thickness scales as E$^{1/2}$ so, at fixed Ekman number, the argument is predicated on the volume of the fluid region increasing, as it does when $d$ is used as the nondimensional lengthscale.
}

{\emph{\hl{Comparison with \textbf{AN-2}. 
}}}
A stark feature of Figure \ref{fig: compare models} is seen in the values for \textbf{AN-2} (used by \cite{dormy2004}), where the trend of Ra$_c$ with $\chi$ reverses compared to the other alternative nondimensionalisations discussed.
\rt{For both heating types the value of Ra$_c$ increases by over a magnitude as the fluid shell becomes thinner.
The key difference of {\bf AN-2} is the choice of lengthscale (now the outer radius of the sphere). Since $r_\text{o}$ is arguably the more appropriate measure for studies focussed on systems with a fixed outer dimension (see discussion in the `Further Remarks' of Section \ref{sec:altform}), the interpretation from Figure \ref{fig: compare models} is that convection is inhibited as the inner core grows.
For thinner shells, the slope of the outer boundary at which the convection columns manifest is steeper, requiring more $z$-dependent motion near the boundaries. 
Since, with $r_\text{o}$ as the nondimensional lengthscale, the total fluid volume decreases as $\chi$ increases, the boundary layers (which have fixed thickness at a given E) also have an enhanced role. This effect is therefore opposite to that described above for {\bf AN-2}.
Hence, a larger level of driving from buoyancy (i.e.,~larger Ra) is needed to overcome the combined effects of the Taylor--Proudman constraint and enhanced influence of the boundary layers at large $\chi$.
The slope at small $\chi$ in the internally heated case is less steep  since the presence of the inner core has a more limited effect until the inner boundary approaches the critical cylinder where the instability is found.
}

The above demonstrates that the choice of nondimensionalisation is clearly important when interpreting results \rt{and comparing the models of different physical systems.}

\section{Discussion}

 \rt{In this section, we address the questions posed in Section \ref{sec:intro} in turn and offer further discussion of our findings.}
 
\emph{\hl{Preference of boundary conditions. 
}}
The transition point \rt{in parameter space} where the \rt{boundary condition that offers the lowest Ra$_c$ changes is found to be}
dependent on $\text{Pr}$, $\chi$, $\text{E}$, and heating type. 
This describes the point at which the Ekman boundary layer \rt{begins to have a} destabilising \rt{effect}.
When varying $\text{Pr}$, 
no-slip boundary conditions 
\rt{give a lower $\text{Ra}_c$ for either $\text{Pr}\gtrsim1.7$ (differential heating case)
or $\text{Pr}\gtrsim0.9$ (internal heating case).}
 When varying $\chi$, the point of transition only \rt{occurs under the internally heated case, and the behaviour depends} on the Ekman number \rt{in a non-trivial fashion}. At $\text{E}=3\times10^{-5}$, the range where no-slip boundary conditions give a higher $\text{Ra}_c$ is $0.47\lesssim\chi\lesssim0.76$ and hence, in this range (and with $\text{Pr}=1$), the Ekman boundary layer is stabilising. However, for a higher Ekman number, $\text{E}=10^{-4}$, there is no range of $\chi$ where this is the case, and for a lower Ekman number, $\text{E}=10^{-5}$, this range of $\chi$ increases.
 \rt{For the {strongly} rotating physical systems motivated by this work, our results suggest that no-slip boundary conditions could be preferred if the shell is thick enough and driven solely by volumetric heat sources (or possibly a combination of internal and differential heating, which is beyond the scope of the current study).
 However, the physical systems typically have low values of Pr, for which our results would suggest a preference for stress-free boundaries. In our work, we have fixed one of either Pr or $\chi$ whilst varying the other. Therefore, further exploration of the parameter space would be needed to determine which boundary condition is more likely to aid convective onset in planetary interiors.   
 }


\emph{\hl{Mixed boundary condition behaviour. 
}} \rt{Under our mixed boundary conditions (with a no-slip inner boundary and a stress-free outer boundary)}, we found that the influence of the outer boundary dominates. Consequently, the critical values of this mixed case are well approximated by the purely stress-free case. This holds when varying either $\text{Pr}$ or $\chi$. In either case, slight variations are found in $\text{Ra}_c$, $m_c$ and $\omega_c$ depending on the effect of the no-slip inner boundary for the heating type, and point in parameter space. \rt{This result suggests that physical systems approximated in mathematical models using a mixed set of conditions (e.g.,~the Sun's convection zone) could also be approximated using purely stress-free conditions, at least near the onset of convection.}

\emph{\hl{Location of convective onset. 
}} 
In the case of differential heating, the critical cylinder (at $s=s_c$) \rt{at which the instability onsets} is found close to the tangent cylinder (i.e.,~$s_c\approx s_L$, as predicted by local theory \cite{roberts1968,busse1970,busse1977}) for all values of $\chi$ \rt{and Pr} tested. 
With internal heating, as Pr is varied, the critical cylinder
\rt{varies in the range $0.54\lesssim s_c/r_\text{o}\lesssim0.69$, a significant departure from the asymptotically predicted value of $s_M\approx0.59r_\text{o}$ for the $\text{Pr}=1$ case \cite{dormy2004}.
Furthermore, the departure from $s_M$ is largest for smaller Ekman numbers with low Pr and low E favouring $s_c<s_M$.
This result suggests that at Ekman and Prandtl numbers applicable to planetary interiors
the convection will onset closer to the inner boundary than predicted for $\text{Pr}=1$ (assuming convection is at least partly driven by volumetric heat sources and $\chi<s_c$). Determining the location more precisely requires further investigations at Ekman numbers lower than those studied in our work.}
\rt{When fixing $\text{Pr}=1$, the expected results of $s_c\approx s_M$} for $\chi\lesssim0.4$ \rt{and $s_c\approx s_L$}, for $\chi\gtrsim0.5$, \rt{were recovered.}
 \rt{For the parameter values tested in our work, $s_c$ appears to undergo a smooth transition from $s_M$ to $s_L$ (as $\chi$ is increased in the interval $0.4<\chi<0.5$).
 Further results within this interval (including at alternative values of E and Pr) would be required to confirm the behaviour.}

{\emph{\hl{Remarks on nondimsionalisation. 
}}}
\rt{A comparison of nondimensionalisations showed that the trend in critical Rayleigh number is dependent on the nondimensional lengthscale chosen. In particular, for a Rayleigh number defined with a fixed outer boundary location (so the total volume of the spherical system is fixed), the onset of convection is inhibited in thinner shells. Conversely, if the volume of the entire system is allowed to vary (by choosing the gap width in the definition of Ra), the larger fluid volume (present even in thinner shells) compensates through additional geometric factors.
Since the former set-up is appropriate for a chosen planet with a growing inner core, future studies focussing on such behaviour could consider using $r_\text{o}$ for nondimensionalisation. This choice has the additional benefit of governing equations that are independent of geometric factors involving $\chi$ (see \hl{Equations }(\ref{f2eqns}a--d)),
with the effect of the aspect ratio appearing through the moving location inner boundary condition with $\chi$. For studies using fixed $\chi$ each nondimensional lengthscale can be regarded as equally informative.}

\emph{\hl{Further studies. 
}}
Additional studies could complement this research. Examining how the radius ratio affects the critical values at different $\text{Pr}$ would help build a more complete understanding of the full parameter space. Investigations into higher and lower values of $\text{Pr}$ and $\chi$ would reveal the effects of different boundary conditions as $\text{Pr}\to0$ and $\chi\to 0 $ or as $\text{Pr}\to\infty$ and $\chi\to 1$. Further work could also clarify our understanding of the transition point where \rt{$1-\text{Ra}^\text{SF}_c/\text{Ra}^\text{NS}_c$ changes sign} and, in particular, the dependence on Ekman number. \rt{Implementing an alternative mixed boundary condition (of a no-slip outer boundary with a stress-free inner boundary)} could also verify that the effect of the outer boundary condition dominates the onset of convection, leading to similar critical values to those of a purely no-slip case. Additionally, a mixture of differential and internal heating types could be explored to determine whether or not the critical values will resemble those of internal or differential heating, analogous to the mixed boundary conditions. The extensive number of critical values across a wide parameter space will serve not only as a basis for further convection studies but also as a valuable resource for simulations of convection-driven dynamos.

\vspace{6pt} 





\authorcontributions{
Conceptualisation, W.S., F.C., R.D.S., and R.J.T.; methodology, W.S., F.C., R.D.S., and R.J.T.; software, R.J.T.; validation, W.S., F.C., and R.J.T.; formal analysis, W.S. and F.C.; investigation, W.S. and F.C.; data curation, W.S. and F.C.; writing---original draft preparation, W.S., F.C., and R.J.T.; writing---review and editing, R.D.S. and R.J.T.; visualisation, W.S. and F.C.; supervision, R.D.S. and R.J.T.; project administration, R.J.T.; funding acquisition, R.D.S. and R.J.T.
All authors have read and agreed to the published version of the manuscript.
}

\funding{\hl{F.C.~is} 
 grateful for an EPSRC PhD scholarship \hl{(EP/W524359/1)}. The work of R.D.S. and R.J.T. was supported by the UK Science and Technology Facilities Council
(grant numbers ST/Y001672/1 and ST/Y00146X/1) . 
}

\dataavailability{
The original data presented in the study are openly available \cite{coke2025}. 

}

\conflictsofinterest{The authors declare no conflicts of interest.
} 





\appendixtitles{no} 
\appendixstart

\appendix




\section[\appendixname~\thesection]{}
\label{Numerics Appendix}
\appendixtitles{yes}

\rt{In this appendix, we further detail our numerical methods including the chosen form for $\mathbf{u}$, the discretisation of the problem, and the form of the matrices $\mathbf{A}$ and $\mathbf{M}$ discussed in Section \ref{sec:methods}.}
\rs{In the following, we adhere closely to the method and choices of \cite{jones2009}.}

\subsection{Poloidal--Toroidal Decomposition}
\label{field decompositions}

\rs{The velocity field is decomposed into poloidal and toroidal parts using scalars $\mathcal{P}$ and $\mathcal{T}$}, such that
\begin{flalign}
    \label{toroidal poloidal decomposition of u}    \mathbf{u}=\nabla\times\nabla\times(r\mathcal{P}\mathbf{\hat{r}})+\nabla\times(r\mathcal{T}\mathbf{\hat{r}}).
\end{flalign}
\hl{This} 
is the most general decomposition that automatically satisfies
the solenoidal condition on $\mathbf{u}$, Equation (\ref{mass equation
  1}). \rs{Taking the $r$-components of the curl and the double curl of
Equation~\eqref{momentum equation 1} yields equations for
$\mathcal{P}$ and $\mathcal{T}$, given by}

\begin{subequations}
\begin{gather}\label{non-ansatz poloidal equation}
    \text{E}\bigg(\frac{\partial}{\partial t}-\nabla^2\bigg)\mathcal{L}^2\nabla^2\mathcal{P}-2\frac{\partial }{\partial\varphi}\nabla^2\mathcal{P}-2\mathcal{Q}\mathcal{T}+\text{ERa}\mathcal{L}^2\Theta=0,\\
\label{non-ansatz toroidal equation}
    \text{E}\bigg(\frac{\partial}{\partial t}-\nabla^2\bigg)\mathcal{L}^2\mathcal{T}-2\frac{\partial \mathcal{T}}{\partial\varphi}+2\mathcal{Q}\mathcal{P}=0,\\
\label{non-ansatz temperature equation}
    \text{Pr}\frac{\partial\Theta}{\partial t}=r^{-1}\Lambda(1-\chi)\Lambda(r)\mathcal{L}^2\mathcal{P}+\nabla^2\Theta.
\end{gather}
\label{TPtheta eqns}
\end{subequations}

\rs{In terms of the poloidal and toroidal scalars, the  impermeable, fixed temperature boundary
  conditions at the inner, $r=r_{\text{i}}$, and outer, $r=r_{\text{o}}$, boundaries become}
\begin{subequations}
\begin{gather}
\label{impermeable boundary condition scalar}
\mathcal{P}= 0\\
\label{theta boundary}
\Theta = 0,
\end{gather}
\rs{and the no-slip and the stress-free boundary conditions become}
\begin{gather}\label{noslip boundaries}
    \frac{\partial \mathcal{P}}{\partial r}=\mathcal{T}=0,\\
\label{stressfree boundaries}
    \frac{\partial^2 \mathcal{P}}{\partial r^2}=\frac{\partial }{\partial r}\bigg(\frac{\mathcal{T}}{r}\bigg)=0,
\end{gather}
\label{TPtheta BCs}
\end{subequations}
respectively.

\subsection{Expansions in Trial Functions and Discretisation}
\label{discretisation}

The following ansatz are then substituted and used for the scalar
fields $\mathcal{P}$, $\mathcal{T}$, and $\Theta$:
\begin{subequations}
\label{governing ansatze}
\begin{gather}
\label{e ansatz}
    \mathcal{T} = \sum^{N_\text{x}+2}_{n=1}\sum^{L}_{l=0}\mathcal{T}_{nl}r^{-1}T_{n}(r)Y^m_{2l+m+1}(\theta,\varphi)e^{\sigma t},\\
\label{f ansatz}
    \mathcal{P} = \sum^{N_\text{x}+4}_{n=1}\sum^{L}_{l=0}\mathcal{P}_{nl}r^{-1}T_{n}(r)Y^m_{2l+m}(\theta,\varphi)e^{\sigma t},\\
 \label{theta ansatz}
    \Theta = \sum^{N_\text{x}+2}_{n=1}\sum^{L}_{l=0}\Theta_{nl}T_{n}(r)Y^m_{2l+m}(\theta,\varphi)e^{\sigma t}.
\end{gather}
\label{TPtheta expansions}
\end{subequations}
\hl{Here} 
, the $T_n(r)$ represents the $n$-th Chebyshev polynomial in the radial coordinate and $Y^m_{\Tilde{l}}(\theta,\varphi)$ is a Schmidt-normalised spherical harmonic \cite{blakely1995}. For each ansatz in (\ref{governing ansatze}), the degree $\Tilde{l}$ is chosen to implement the Busse symmetry \cite{busse1970,busse1982} found at convective onset. $N_\text{x}$ and $L$ give the radial and angular resolutions, respectively, while any additional terms appearing in the radial summation account for the imposed boundary conditions on that scalar field.

\subsection{Projection onto the Trial Functions and Matrix Expressions}
\label{matrices}

\textls[-15]{By substituting the forms given by Equation (\ref{TPtheta expansions}a--c) into Equations (\ref{TPtheta eqns}a--c) and~\mbox{(\ref{TPtheta BCs}a--d)}}, the matrices $\mathbf{A}$ and $\mathbf{M}$ that form the eigenvalue problem of Equation (\ref{eigenvalue problem1}) can be constructed. The time dependence will enter through the eigenvalue, $\sigma$, and so all terms with a factor of $\sigma$ are entered into $\mathbf{M}$. The remaining terms make up the $\mathbf{A}$ matrix. With these equations, we can outline the explicit forms of Equations (\ref{TPtheta eqns}a--c) and~(\ref{TPtheta BCs}a--d) in terms of the Fourier coefficients $\mathcal{P}_{nl}$, $\mathcal{T}_{nl}$, and $\Theta_{nl}$.
Note that, in what follows, we use the notation $T^{(i)}_n(r) = \frac{d^i T_n}{d r^i}(r)$.

Starting with the poloidal equation, Equation\,(\ref{non-ansatz poloidal equation}), we obtain:
\begin{subequations}
\begin{flalign}
\label{eigenvalue problem f}
        \sigma \text{E}\sum^{N_{\text{x}}+4}_{n=1}\sum^{L}_{l=0}\mathcal{P}_{nl}\lambda_0 (\lambda_0 + 1)&\bigg[T^{(2)}_{n}-\lambda_0 (\lambda_0 + 1)r^{-2}T_{n}\bigg]=\text{E}\sum^{N_{\text{x}}+4}_{n=1}\sum^{L}_{l=0}\mathcal{P}_{nl}\lambda_0 (\lambda_0 + 1)\bigg[T^{(4)}_{n}-2\lambda_0 (\lambda_0 + 1)r^{-2}T^{(2)}_{n}\nonumber\\
        &+4\lambda_0 (\lambda_0 + 1)r^{-3}T^{(1)}_{n}+\lambda_0 (\lambda_0 + 1)(\lambda_0 (\lambda_0 + 1)-6)r^{-4}T_{n}\bigg]\nonumber\\
        &+2im\sum^{N_{\text{x}}+4}_{n=1}\sum^{L}_{l=0}\mathcal{P}_{nl}\bigg[T^{(2)}_{n}-\lambda_0 (\lambda_0 + 1)r^{-2}T_{n}\bigg]\nonumber\\
        &-2\sum^{N_{\text{x}}+2}_{n=1}\sum^{L}_{l=0}\mathcal{T}_{nl}\frac{\lambda_0(\lambda_0+2)}{2\lambda_0+3}\sqrt{(\lambda_0 +1)^2-m^2}\bigg[T^{(1)}_{n}+(\lambda_0+1)r^{-1}T_{n}\bigg]\nonumber\\
        &-2\sum^{N_{\text{x}}+2}_{n=1}\sum^{L+1}_{l=1}\mathcal{T}_{nl}\frac{\lambda_0^2-1}{2\lambda_0-1}\sqrt{\lambda_0^2-m^2}\bigg[T^{(1)}_{n}-\lambda_0 r^{-1}T_{n}\bigg]\nonumber\\
        &-\text{ERa}\sum^{N_{\text{x}}+2}_{n=1}\sum^{L}_{l=0}\Theta_{nl}\lambda_0 (\lambda_0 + 1)rT_{n},
\end{flalign}
where $\lambda_0 = 2l+m$. The toroidal equation, Equation\,(\ref{non-ansatz toroidal equation}), has a similar form: \vspace{-6pt}
\begin{flalign}
\label{eigenvalue problem e}
        \sigma \text{E}\sum^{N_{\text{x}}+2}_{n=1}\sum^{L}_{l=0}\mathcal{T}_{nl}\lambda_1(\lambda_1 +1)T_{n}&=2im\sum^{N_{\text{x}}+2}_{n=1}\sum^{L}_{l=0}\mathcal{T}_{nl}T_{n}\nonumber\\
        &\qquad+\text{E}\sum^{N_{\text{x}}+2}_{n=1}\sum^{L}_{l=0}\mathcal{T}_{nl}\lambda_1(\lambda_1 +1)\bigg[T^{(2)}_{n}-\lambda_1(\lambda_1 +1)r^{-2}T_{n}\bigg]\nonumber\\
        &\qquad+2\sum^{N_{\text{x}}+4}_{n=1}\sum^{L-1}_{l=-1}\mathcal{P}_{nl}\frac{\lambda_1(\lambda_1+2)}{2\lambda_1 +3}\sqrt{(\lambda_1+1)^2-m^2}\bigg[T^{(1)}_{n}+(\lambda_1+1)r^{-1}T_{n}\bigg]\nonumber\\
        &\qquad+2\sum^{N_{\text{x}}+4}_{n=1}\sum^{L}_{l=0}\mathcal{P}_{nl}\frac{\lambda_1^2-1}{2\lambda_1-1}\sqrt{\lambda_1^2-m^2}\bigg[T^{(1)}_{n}-\lambda_1 r^{-1}T_{n}\bigg],
\end{flalign}
where $\lambda_1=2l+m+1$. Lastly, the heat equation, Equation\,(\ref{non-ansatz temperature equation}), gives:
\begin{flalign}
\label{eigenvalue problem theta}
        \sigma \text{Pr}\sum^{N_{\text{x}}+2}_{n=1}\sum^{L}_{l=0}\Theta_{nl}T_{n}&=\sum^{N_{\text{x}}+4}_{n=1}\sum^{L}_{l=0}\mathcal{P}_{nl}\lambda_0(\lambda_0+1)r^{-2}\lambda_0(1-\chi)\lambda_0(r)T_{n}\nonumber\\
        &\qquad+\sum^{N_{\text{x}}+2}_{n=1}\sum^{L}_{l=0}\Theta_{nl}\bigg[T^{(2)}_{n}+2r^{-1}T^{(1)}_{n}-\lambda_0(\lambda_0+1)r^{-2}T_{n}\bigg],
\end{flalign}
where, again, $\lambda_0 = 2l+m$. 
\end{subequations}

\begin{subequations}  
The boundary value problem is completed by the boundary conditions found in Equation (\ref{TPtheta BCs}a--d). Applying the ansatz to these conditions, we obtain the following form for each type of conditions. The impermeable boundary condition leads to
\begin{flalign}
    0=\sum^{N_{\text{x}}+4}_{n=1}\sum^{L}_{l=0}\mathcal{P}_{nl}T_{n}(r),
\end{flalign}
while the fixed temperature boundary imposes
\begin{flalign}
    0=\sum^{N_{\text{x}}+2}_{n=1}\sum^{L}_{l=0}\Theta_{nl}T_n(r).
\end{flalign}

The no-slip conditions become
\begin{flalign}
0 = \sum^{N_{\text{x}}+4}_{n=1} \sum^{L}_{l=0} \mathcal{P}_{nl} T^{(1)}_{n}(r)
\qquad \text{and} \qquad
0 = \sum^{N_{\text{x}}+2}_{n=1} \sum^{L}_{l=0} \mathcal{T}_{nl} T_{n}(r),
\end{flalign}
while the stress-free conditions become 
\begin{flalign}
    \label{stress-free expansion ansatz}0=\sum^{N_{\text{x}}+4}_{n=1}\sum^{L}_{l=0}\mathcal{P}_{nl}\bigg[rT^{(2)}_{n}(r)-2T^{(1)}_n(r)\bigg]
    \qquad \text{and} \qquad
    0=\sum^{N_\text{x}+2}_{n=1}\sum^{L}_{l=0}\mathcal{T}_{nl}\bigg[rT^{(1)}_n-2T_n\bigg].
\end{flalign}
\end{subequations}

Finally, the eigenvector $\mathbf{v}$ appearing in Equation (\ref{eigenvalue problem1}) is composed of the  coefficients $\mathcal{P}_{nl},\ \mathcal{T}_{nl}$, and $\Theta_{nl}$. For an eigenvector with definition $N_\text{x}$ and $L$, it is structured as follows:
\begin{align}
    \mathbf{v}  =
    &\Big[
        \mathcal{P}_{10} ~\dots~ \mathcal{P}_{(N_\text{x}+4)0} ~~~~
        \mathcal{T}_{10} ~\dots~ \mathcal{T}_{(N_\text{x}+2)0} ~~~~
        \Theta_{10}  ~\dots~ \Theta_{(N_\text{x}+2)0} \nonumber \\
       & ~\dots~\mathcal{P}_{1L}  ~\dots~ \mathcal{P}_{(N_\text{x}+4)L}
        ~~~~ \mathcal{T}_{1L}  ~\dots~ \mathcal{T}_{(N_\text{x}+2)L} ~~~~
        \Theta_{1L}  ~\dots~ \Theta_{(N_\text{x}+2)L} \Big]^T,
    \label{eigenvector v form}
\end{align}
and has length $(3(N_\text{x}+2)+2)(L+1)$ to account for each value of $n$ and $l$. Once $\mathbf{v}$ and, by extension, $\mathcal{P}_{nl},\ \mathcal{T}_{nl}$, and $\Theta_{nl}$ are determined, the temperature field can be constructed via Equation (\ref{theta ansatz}), and the velocity field is determined from Equations (\ref{toroidal poloidal decomposition of u}), (\ref{e ansatz}), and (\ref{f ansatz}).

\appendixtitles{no}
\section[]{}
\label{Convergence Test}

\begin{figure}[H]
\subfloat[\centering]{\includegraphics[width=0.48\linewidth]{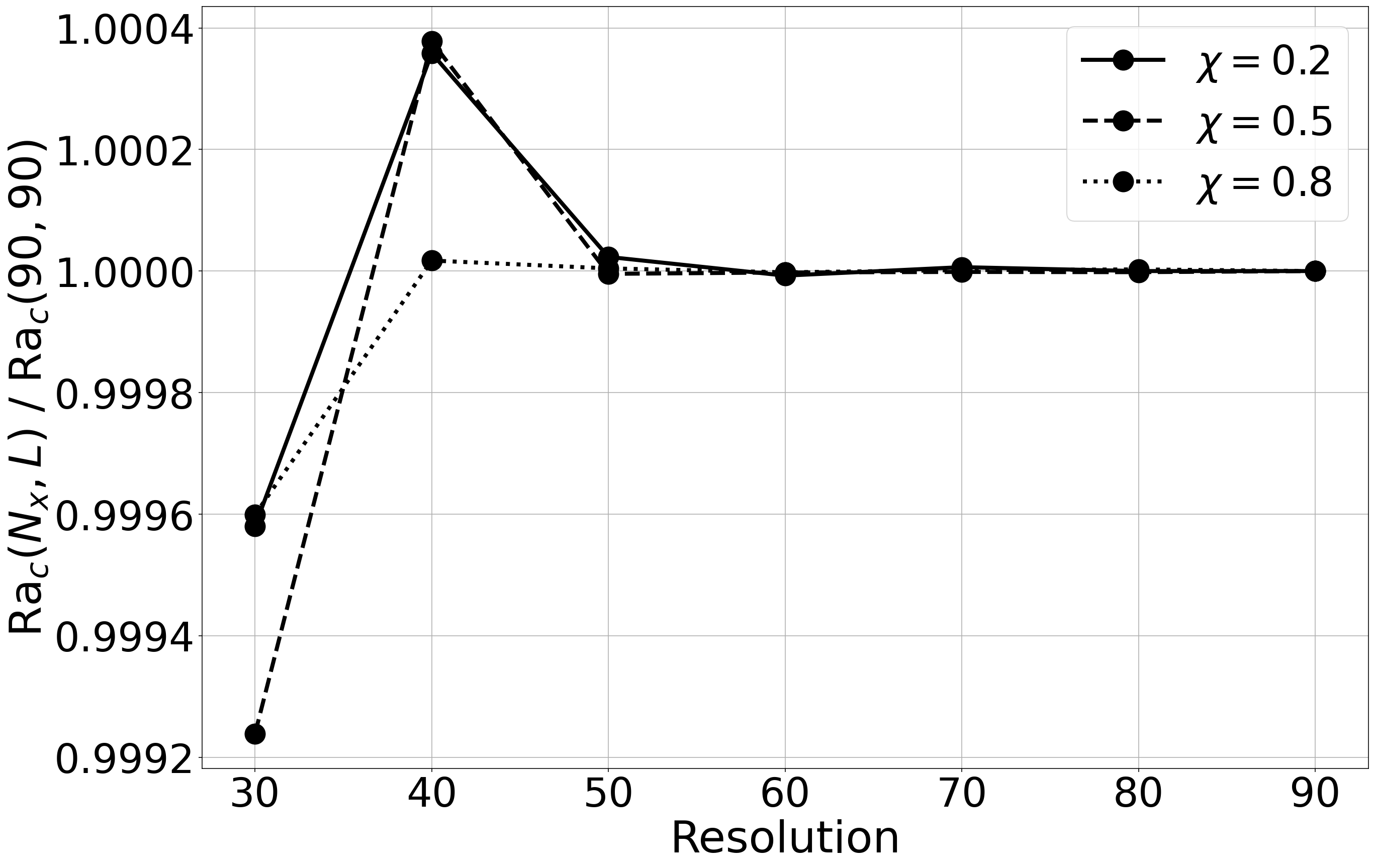}}
\subfloat[\centering]{\includegraphics[width=0.48\linewidth]{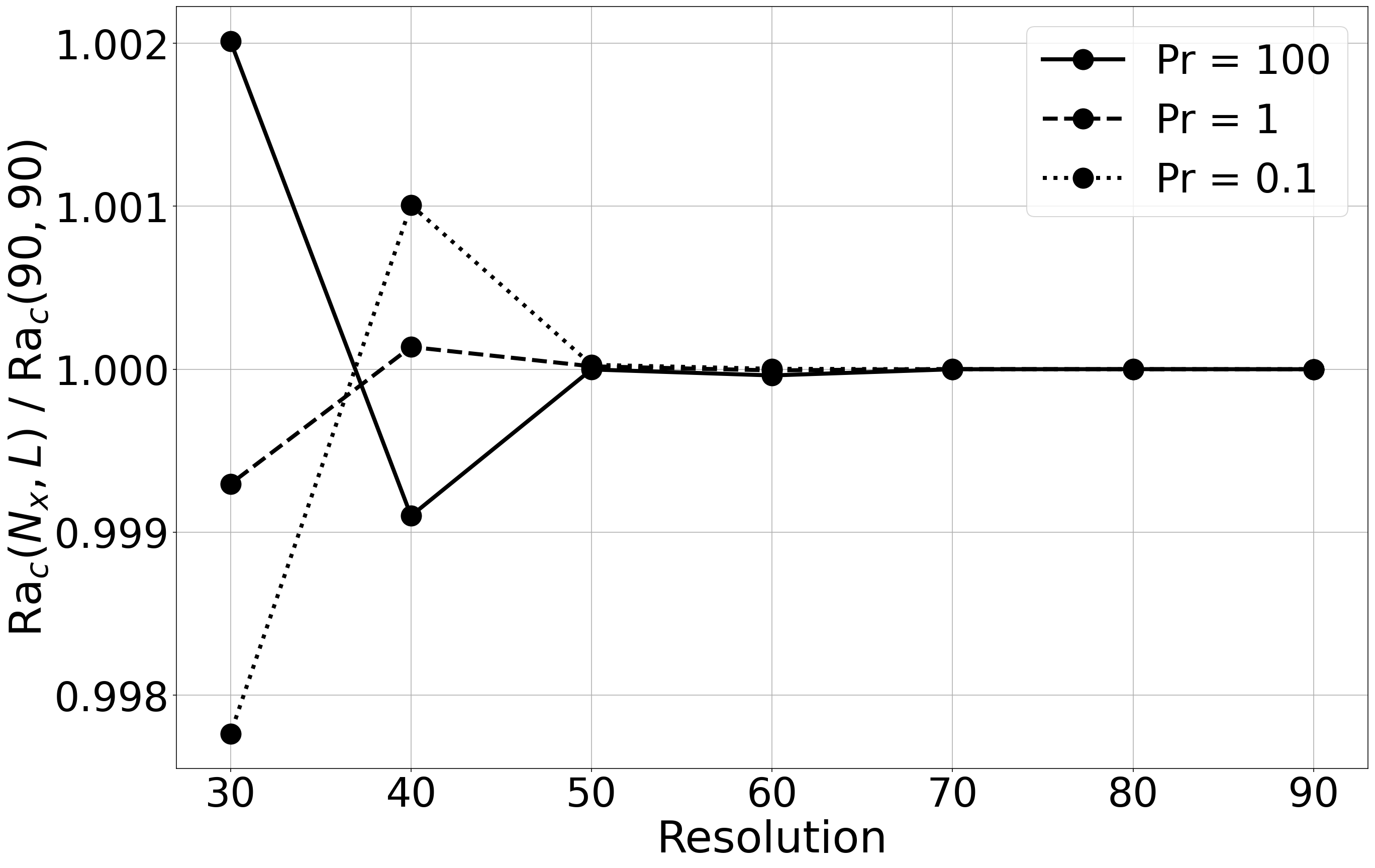}}
\caption{{Dependence of $\text{Ra}_c$ on resolution for (\textbf{a}) different values of $\chi$ with $\text{Pr}=1$ and (\textbf{b}) different values of Pr with $\chi=0.35$. Differential heating and purely no-slip boundary conditions are applied and $\text{E}=10^{-5}$. As $N_\text{x}=L$, the horizontal axis represents both resolution parameters. }}
\label{Res Tests}
\end{figure}

{In this appendix, we detail convergence tests performed to validate our choice of resolution. For $\text{E}=10^{-5}$ (the most numerically demanding case studied), we consider $\text{Ra}_c(N_x,L)$ as a function of the resolution parameters, $N_x$ and $L$, with $N_x=L$.}
{Figure \ref{Res Tests} shows that} {critical Rayleigh numbers {have converged to} within $0.003\%$ of $\text{Ra}_c$ at $N_\text{x}=L=90$ when a resolution of $N_\text{x}=L\ge 50$ is used. {This is the case for} Pr and $\chi$ values {tested from across the full range of values used in this study}. 
}

\reftitle{References}

\PublishersNote{}
\end{document}